\documentclass[extra, mreferee]{gji}

\usepackage{epsfig, color}
\usepackage{amssymb, amsmath}
\usepackage{setspace}

\title[Propagator and transfer matrices]
 {Propagator and transfer matrices, Marchenko focusing functions and their mutual relations}

\author[Wapenaar et al.]
{\small Kees Wapenaar$^1$, Marcin Dukalski$^2$, Christian Reinicke$^2$ and Roel Snieder$^3$\\
$^1$Department of Geoscience and Engineering, Delft University of Technology, The Netherlands\\
$^2$Aramco Delft Global Research Center, The Netherlands\\
$^3$Center for Wave Phenomena, Colorado School of Mines, Golden CO 80401, USA}
\pagerange{1--99}
\volume{999}
\pubyear{2023}

\begin{document}
\begin{spacing}{2.0}
\label{firstpage}

\maketitle

\begin{summary}
{\small 
Many seismic imaging methods use wave field extrapolation operators to redatum sources and receivers from the surface into the subsurface. 
We discuss wave field extrapolation operators that account for internal multiple reflections, in particular propagator matrices, transfer matrices and Marchenko focusing functions.

A propagator matrix is a square matrix that `propagates' a wave-field vector from one depth level to another. 
It accounts for primaries and multiples and holds for propagating and evanescent waves. 

A Marchenko focusing function is a wave field that focuses at a designated point in space at zero time.
Marchenko focusing functions are useful for retrieving the wave field inside a heterogeneous medium from the reflection response at its surface. By expressing these focusing functions
in terms of the propagator matrix, the usual approximations (such as ignoring evanescent waves) are avoided.

While a propagator matrix acts on the full wave-field vector, a transfer matrix (according to the definition employed in this paper)
`transfers' a decomposed wave-field vector (containing downgoing and upgoing waves) from one depth level to another.
It can be expressed in terms of decomposed Marchenko focusing functions.

We present propagator matrices, transfer matrices and Marchenko focusing functions in a consistent way and discuss their mutual relations.
In the main text we consider the acoustic situation and in the appendices we discuss other wave phenomena.
Understanding these mutual connections may lead to new developments of Marchenko theory and its applications in wave field focusing, Green's function retrieval and imaging.}
\end{summary}


\section{Introduction}\label{sec1}

In many seismic imaging methods, wave field extrapolation is used to redatum sources and receivers from the surface to a depth level in the subsurface.
In most cases the redatuming process is based on one-way wave field extrapolation operators, which only account for primaries. 
To account for internal multiple reflections in redatuming, more advanced wave field extrapolation operators are required. 
This paper is not about the redatuming process itself, but about wave field extrapolation operators that account for internal multiples. 
In particular, we discuss propagator and transfer matrices, Marchenko focusing functions and their mutual relations.

In elastodynamic wave theory, a propagator matrix is a square matrix that `propagates' a wave-field vector from one depth level to another. 
It was originally introduced in geophysics for horizontally layered media \citep{Thomson50JAP, Haskell53BSSA, Gilbert66GEO} and
later extended for laterally varying media \citep{Kennett72GJRAS}. It has been used for modelling surface waves \citep{Woodhouse74GJR}
and reflection and transmission responses of heterogeneous media \citep{Haines88GJI, Kennett90GJI, Koketsu91GJI, Takenaka93WM}.
It has also been proposed as an operator for accurate seismic imaging schemes, accounting for high propagation angles \citep{Kosloff83GEO} and internal multiple reflections \citep{Wapenaar86GP2}.
The wave-field vector that the propagator matrix acts on contains components of the full wave field (e.g. particle velocity and stress). 
Here `full' means that the wave field implicitly consists of downgoing and upgoing, propagating and evanescent waves.

A Marchenko focusing function is a wave field that focuses at a designated point in space at zero time, accounting for primaries and multiples.
Marchenko focusing functions were originally introduced to retrieve the wave field inside a horizontally layered medium from the reflection response at the 
boundary of that medium \citep{Rose2001PRA, Rose2002IP, Broggini2012EJP, Slob2014GEO}. This has been extended for laterally varying media \citep{Wapenaar2013PRL},
under the assumption that the wave field inside the medium can be decomposed into downgoing and upgoing components and that evanescent waves can be neglected. 
It has recently been shown that the propagator matrix can be expressed in terms of Marchenko focusing functions and vice versa \citep{Wapenaar2022GEO}.
Via this relation, the usual assumptions underlying the focusing functions (such as ignoring evanescent waves) are circumvented.

In this paper we define a transfer matrix as a square matrix that `transfers' decomposed wave-field vectors (explicitly containing downgoing and upgoing waves) from one depth level to another \citep{Born65Book, Katsidis2002AO}. 
It is different from the propagator matrix, which acts on full wave-field vectors
(but please note that in the literature there is not a clear distinction between the use of the terminologies `propagator matrix' and `transfer matrix'). It has recently been shown that 
the transfer matrix can be expressed in terms of decomposed Marchenko focusing functions 
\citep{Dukalski2022EAGE, Dukalski2022IMAGE}, an insight that is expected to be useful in further analysis of the minimum-phase property of elastodynamic focusing functions \citep{Reinicke2023GJI} 
and beyond.

The aim of this paper is to present  propagator matrices, transfer matrices and Marchenko focusing functions in a consistent way and to discuss their mutual relations.
We aim to set up the theory as general as possible, accounting for lateral and vertical variations of the medium parameters, accounting for evanescent waves and taking dissipation into account.
Whereas in the main text we consider acoustic waves, in the appendices we generalize the theory for other wave phenomena.
The numerical examples, which are meant as illustrations of the different quantities and their relations, 
are restricted to oblique acoustic plane waves in a lossless horizontally layered medium. 

We hope that this consistent treatment will contribute to the understanding of the mutual connections and provide insight in the assumptions and approximations that underlie 
Marchenko-type wave field retrieval schemes and how to cope with them
\citep{Slob2016PRL, Dukalski2019GJI, Dukalski2022IMAGE, Reinicke2020GEO, Reinicke2023GJI, Elison2020GJI, Diekmann2021PRR, Wapenaar2021GJI,  Kiraz2023GJI}.
Moreover, we hope to stimulate new research directions.

The setup of this paper is as follows. In section \ref{sec2} we discuss the $2\times 2$ propagator matrix for acoustic wave fields and its relation with acoustic Marchenko focusing functions.
The advantage of concentrating on the acoustic situation is that all expressions are relatively simple and yet contain all essential aspects. 
In section \ref{sec3} we discuss the $2\times 2$ transfer matrix for acoustic wave fields and its relation with decomposed acoustic Marchenko focusing functions. 
In section \ref{sec6} we present some conclusions.

Appendices \ref{AppA} and \ref{AppB} are generalisations of sections \ref{sec2} and \ref{sec3} for other wave phenomena. 
Here, the propagator and transfer matrices are $N\times N$ matrices, with $N$ ranging from 2 for acoustic waves to 12 for seismoelectric waves; 
the Marchenko focusing functions are $\frac{N}{2}\times\frac{N}{2}$ matrices.
We derive their mutual relations by exploiting general symmetry properties, which are derived in Appendix \ref{AppC}.
The appendices not only cover classical waves, but also quantum mechanical waves obeying the Schr\"odinger equation ($N=2$) and the Dirac equation ($N=4$).

\section{Acoustic propagator matrix and focusing functions}\label{sec2}

\subsection{Acoustic matrix-vector wave equation}

Our starting point is the following matrix-vector wave equation in the space-frequency domain 
\begin{eqnarray}\label{eq2.0}
\partial_3{\bf q} = {{\mbox{\boldmath ${\cal A}$}}}\,{\bf q} +{\bf d}
\end{eqnarray}
\citep{Woodhouse74GJR, Corones75JMAA, Ursin83GEO, Kosloff83GEO, Fishman84JMP, Wapenaar86GP2, Hoop96JMP}.
In Appendix \ref{AppA} we discuss this equation for a range of wave phenomena. Here we consider acoustic waves. For this situation,
 ${\bf q}$ is a vector containing the wave field components $p$ (acoustic pressure) and $v_3$ (vertical component of the particle velocity), both as a function
of the space coordinate vector ${\bf x}=(x_1,x_2,x_3)$ (with positive $x_3$ denoting depth) and the angular frequency $\omega$, hence,
\begin{eqnarray}\label{eq9996ge}
{\bf q}({\bf x},\omega) = \begin{pmatrix} p \\ v_3 \end{pmatrix}({\bf x},\omega).
\end{eqnarray}
Operator $\partial_3$ stands for the partial differential operator $\partial/\partial x_3$.
The space- and frequency-dependent operator matrix ${{{\mbox{\boldmath ${\cal A}$}}}}$ is defined as
\begin{eqnarray}
{{\mbox{\boldmath ${\cal A}$}}}({\bf x},\omega)&=& 
 \begin{pmatrix} 0     & i\omega \rho \\
i\omega\kappa-\frac{1}{i\omega}\partial_\alpha\frac{1}{\rho}\partial_\alpha & 0  \end{pmatrix}({\bf x},\omega),
\end{eqnarray}
where $\kappa({\bf x},\omega)$ is the compressibility, $\rho({\bf x},\omega)$ the mass density and $i$ the imaginary unit. 
Operator $\partial_\alpha$ stands for the partial differential operator $\partial/\partial x_\alpha$.
Greek subscripts take on the values 1 and 2 and Einstein's summation convention applies to repeated subscripts, unless otherwise noted.
In general the medium may be dissipative,
meaning that $\kappa$ and $\rho$ may be frequency-dependent and complex-valued, with (for positive $\omega$) $\Im(\kappa)\ge 0$ and $\Im(\rho)\ge 0$, 
where $\Im$ denotes the imaginary part. 
For later convenience we rewrite the operator matrix as follows
\begin{eqnarray}
{{\mbox{\boldmath ${\cal A}$}}}({\bf x},\omega)
&=&\begin{pmatrix} 0     & i\omega \rho \\
-\frac{1}{i\omega\sqrt{\rho}}{\cal H}_2\frac{1}{\sqrt{\rho}} & 0  \end{pmatrix}({\bf x},\omega).\label{eqAcoustic}
\end{eqnarray}
Here ${\cal H}_2({\bf x},\omega)$ is the Helmholtz operator, defined as
\begin{eqnarray}
{\cal H}_2({\bf x},\omega)=k^2({\bf x},\omega)+\partial_\alpha\partial_\alpha, \label{eqHelmholtz}
\end{eqnarray}
with wavenumber $k({\bf x},\omega)$ defined via
\begin{eqnarray}
k^2({\bf x},\omega)=\omega^2\kappa\rho-\frac{3(\partial_\alpha\rho)(\partial_\alpha\rho)}{4\rho^2}+\frac{(\partial_\alpha\partial_\alpha\rho)}{2\rho}\label{eqks}
\end{eqnarray}
\citep{Brekhovskikh60Book, Wapenaar2001RS}.
Finally, vector ${\bf d}$ in equation (\ref{eq2.0}) contains source terms, according to
\begin{eqnarray}\label{eq9996ged}
 {\bf d}({\bf x},\omega) = \begin{pmatrix} \hat f_3 \\ \frac{1}{i\omega}\partial_\alpha(\frac{1}{\rho}\hat f_\alpha) + \hat q \end{pmatrix}({\bf x},\omega).
\end{eqnarray}
Here $\hat f_\alpha({\bf x},\omega)$ and $\hat f_3({\bf x},\omega)$ are the horizontal and vertical components, respectively, of the external force density 
(the hats are used to distinguish external force components from focusing functions),
and $\hat q({\bf x},\omega)$ is the volume injection-rate density (where $\hat q$
is to be distinguished from the wave field vector ${\bf q}$). From here onward we simplify the notation by not explicitly mentioning the frequency-dependency in the argument lists.

\subsection{Acoustic propagator matrix}

\begin{figure}
\centerline{\epsfysize=9. cm \epsfbox{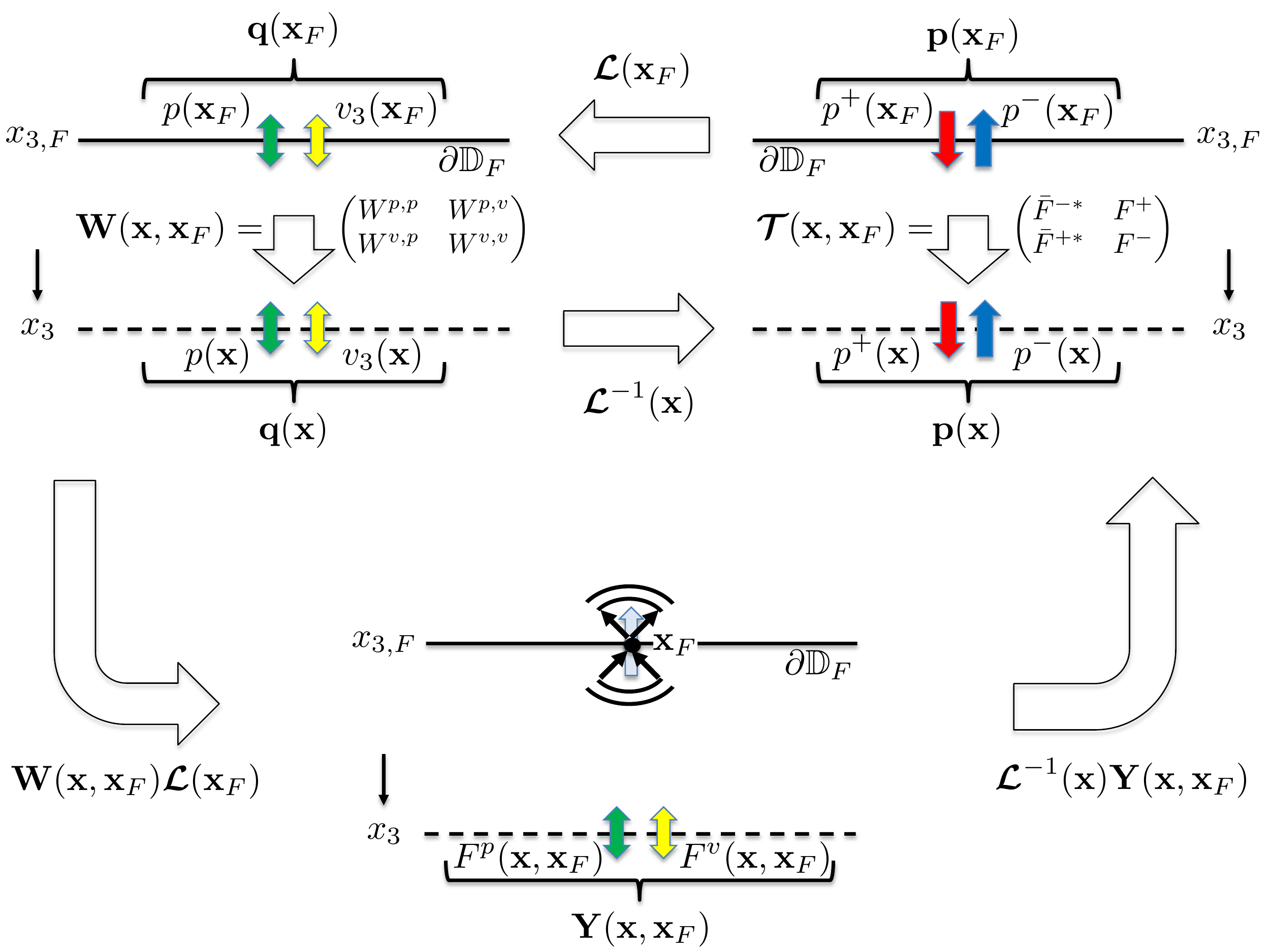}}
\caption{Relations between the propagator matrix ${\bf W}({\bf x},{\bf x}_F)$, the transfer matrix ${{{\mbox{\boldmath ${\cal T}$}}}}({\bf x},{\bf x}_F)$, 
and the Marchenko focusing functions $F^p({\bf x},{\bf x}_F)$ and
$F^v({\bf x},{\bf x}_F)$ (right column of ${\bf Y}({\bf x},{\bf x}_F)$). 
The green and yellow double-sided arrows indicate full wave fields (implicitly consisting of downgoing and upgoing components), whereas the red and blue
single-sided arrows indicate decomposed downgoing and upgoing wave fields, respectively.}\label{Fig1}
\end{figure}

We define a boundary $\partial\mathbb{D}_F$ at depth level $x_3=x_{3,F}$. We define a coordinate vector ${\bf x}_F$ at this boundary as  ${\bf x}_F=(x_{1,F},x_{2,F},x_{3,F})$ (with fixed $x_{3,F}$).
We introduce the propagator matrix ${\bf W}({\bf x},{\bf x}_F)$ as a solution of wave equation (\ref{eq2.0}) for the source-free situation, according to
\begin{eqnarray}\label{eq2.1}
\partial_3{\bf W}({\bf x},{\bf x}_F) = {{\mbox{\boldmath ${\cal A}$}}}({\bf x}){\bf W}({\bf x},{\bf x}_F),
\end{eqnarray}
with boundary condition
\begin{eqnarray}
{\bf W}({\bf x},{\bf x}_F)|_{x_3={x_{3,F}}} = {\bf I}\delta({{\bf x}_{\rm H}}-{{\bf x}_{{\rm H},F}}),\label{eq9998d}
\end{eqnarray}
where ${\bf I}$ is the identity matrix and ${\bf x}_{\rm H}$ and ${\bf x}_{{\rm H},F}$ denote the horizontal coordinates of ${\bf x}$ and ${\bf x}_F$, respectively, hence
${\bf x}_{\rm H}=(x_1,x_2)$ and ${\bf x}_{{\rm H},F}=(x_{1,F},x_{2,F})$. Since equations (\ref{eq2.0}) and (\ref{eq2.1}) are both linear, Huygens' superposition principle can be applied
to get a representation for ${\bf q}({\bf x})$ in terms of ${\bf W}({\bf x},{\bf x}_F)$. For a given depth level $x_3$, assuming there are no sources for ${\bf q}({\bf x})$ between $x_{3,F}$ and $x_3$, 
we obtain
\begin{eqnarray}\label{eq1330}
{\bf q}({\bf x})&=&\int_{\partial\mathbb{D}_F} {\bf W}({\bf x},{\bf x}_F){\bf q}({\bf x}_F){\rm d}^2{\bf x}_F
\end{eqnarray}
\citep{Gilbert66GEO, Kennett72GJRAS, Woodhouse74GJR}. 
Note that equation (\ref{eq1330}) expresses the `propagation' of ${\bf q}$ from depth level $x_{3,F}$ to depth level $x_3$, which is why ${\bf W}({\bf x},{\bf x}_F)$ is called the propagator matrix.
It is partitioned as follows
\begin{eqnarray}\label{eq424}
{\bf W}({\bf x},{\bf x}_F)= \begin{pmatrix}W^{p,p}      & W^{p,v} \\
                W^{v,p} & W^{v,v}    \end{pmatrix}({\bf x},{\bf x}_F),
\end{eqnarray}
where $W^{p,p}$, $W^{p,v}$, $W^{v,p}$ and $W^{v,v}$ are the scalar components of the propagator matrix. 
For each of these components, the second superscript refers to the wave field component ($p$ or $v_3$) it acts on at ${\bf x}_F$, 
whereas the first superscript refers to the wave field component it contributes to at ${\bf x}$.
Equation (\ref{eq1330}) is illustrated in the upper-left frame of Figure \ref{Fig1}. The solid line at $x_{3,F}$ denotes the boundary $\partial\mathbb{D}_F$ (not necessarily a physical boundary). 
The medium below $\partial\mathbb{D}_F$ may be inhomogeneous and dissipative. The dashed line at $x_3$ indicates an arbitrary depth level inside the inhomogeneous medium.

By applying equation (\ref{eq1330}) recursively, it follows that ${\bf W}$ obeys the following recursive expression
\begin{eqnarray}\label{eq1330k}
{\bf W}({\bf x}',{\bf x}_F)&=&\int_{{{\partial\mathbb{D}}}}{\bf W}({\bf x}',{\bf x}){\bf W}({\bf x},{\bf x}_F){\rm d}^2{\bf x},
\end{eqnarray}
where ${{{\partial\mathbb{D}}}}$ is a horizontal boundary at a constant depth level $x_3$. By taking $x_3'=x_{3,F}$, we obtain from equations (\ref{eq9998d}) and (\ref{eq1330k}) 
\begin{eqnarray}\label{eq1330kinv}
{\bf I}\delta({{\bf x}_{\rm H}'}-{{\bf x}_{{\rm H},F}})&=&\int_{{{\partial\mathbb{D}}}}{\bf W}({\bf x}_{\rm H}',x_{3,F},{\bf x}){\bf W}({\bf x},{\bf x}_F){\rm d}^2{\bf x},
\end{eqnarray}
from which it follows that ${\bf W}({\bf x}_F,{\bf x})$ is the inverse of ${\bf W}({\bf x},{\bf x}_F)$.

The propagator matrix ${\bf W}({\bf x},{\bf x}_F)$ accounts for primaries and multiples between $x_{3,F}$ and $x_3$ and it holds for propagating and evanescent waves 
(for example, \cite{Woodhouse74GJR} uses the elastodynamic version of the propagator matrix to analyse surface waves).
Evanescent field components may lead to instability and should be handled with care \citep{Kennett79GJRAS}.
Since the underlying wave equation is based on the explicit Helmholtz operator ${\cal H}_2$ (rather than on its square-root, appearing in one-way wave equations),
\cite{Kosloff83GEO} argue that the numerical evaluation of equation (\ref{eq1330}) converges much faster and for higher propagation angles than schemes based on one-way wave equations. 
They exploit this property in wide-angle imaging of seismic reflection responses. They use filters to eliminate evanescent and downgoing waves, so they do not exploit the
fact that the propagator matrix can handle multiply reflected and evanescent waves. 
\cite{Wapenaar86GP2} propose a seismic imaging scheme based on the propagator matrix that handles internal multiple reflections.
Since their scheme is very sensitive to the chosen background model it has not found broad applications. 
In section \ref{sec2.3} we show that the propagator matrix can be expressed in terms 
of Marchenko focusing functions. For a lossless medium, these focusing functions can be derived from seismic reflection data and a smooth background model (section \ref{sec2.4}). 
Hence, this leads to a propagator matrix that can be used for seismic imaging, which properly handles internal multiple reflections without being highly sensitive to the background model.

\begin{figure}
\centerline{\epsfysize=7.cm \epsfbox{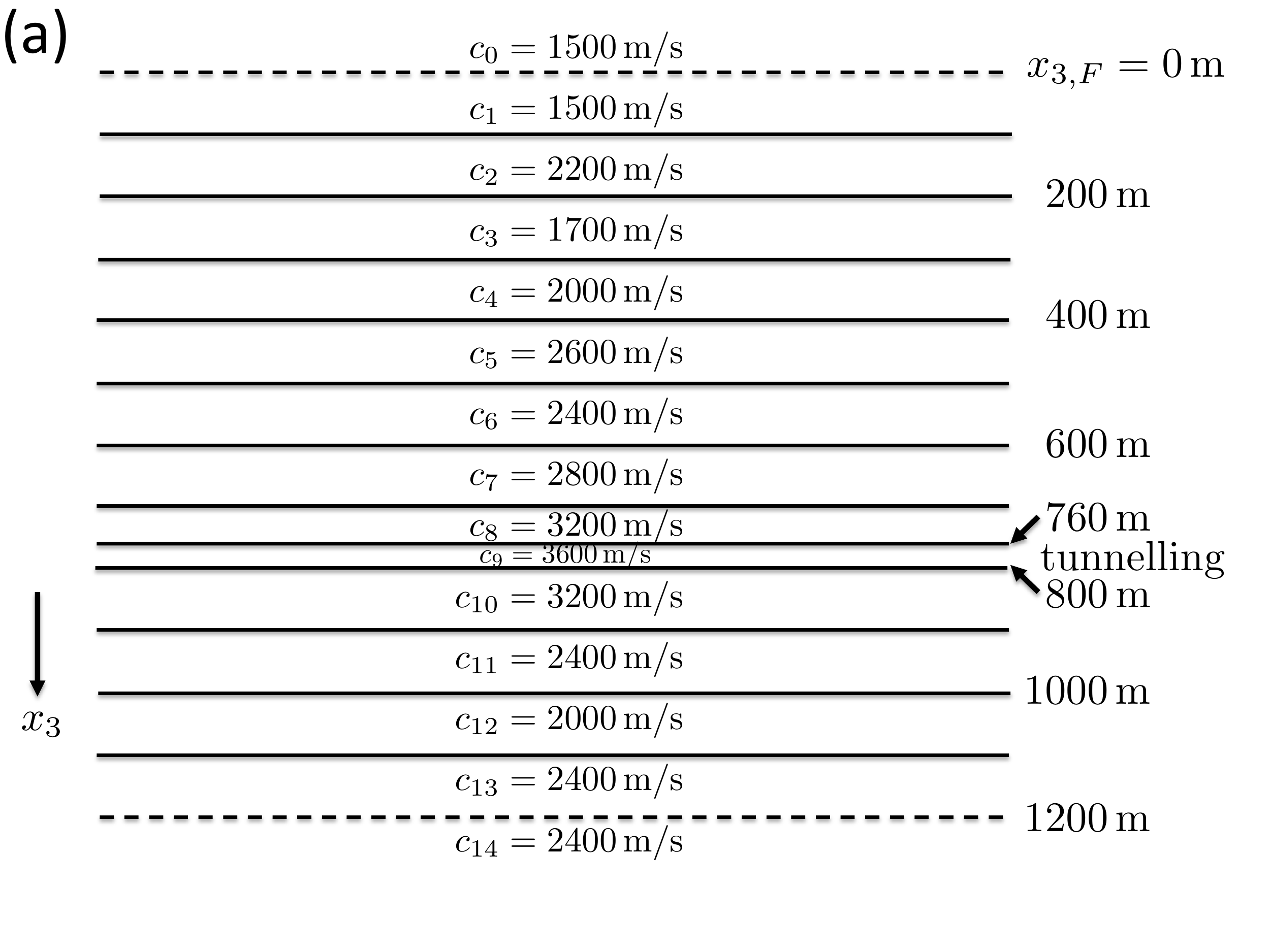}}
\vspace{0.cm}
\centerline{\epsfysize=7. cm \epsfbox{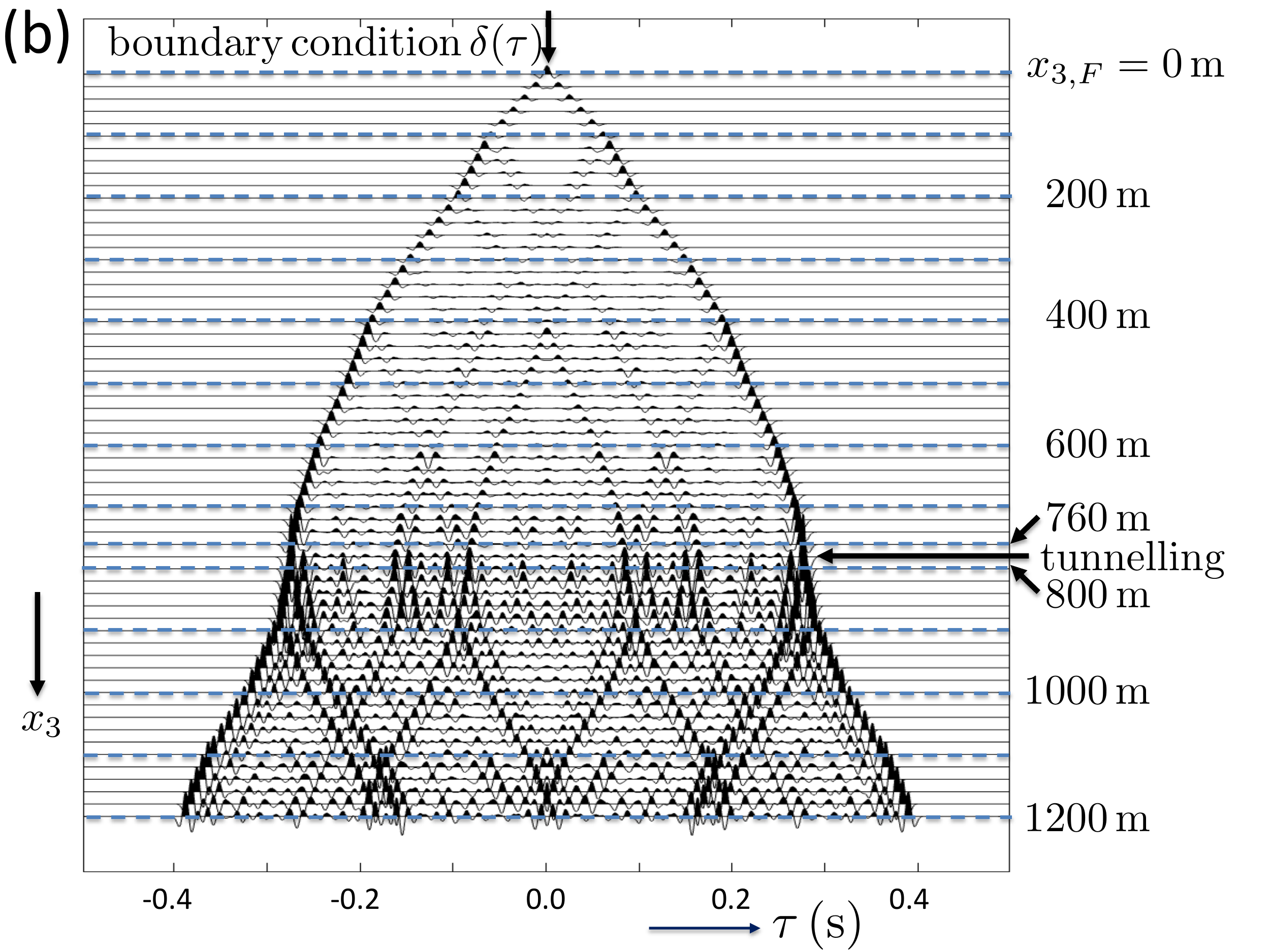}}
\vspace{0.4cm}
\centerline{\epsfysize=7. cm \epsfbox{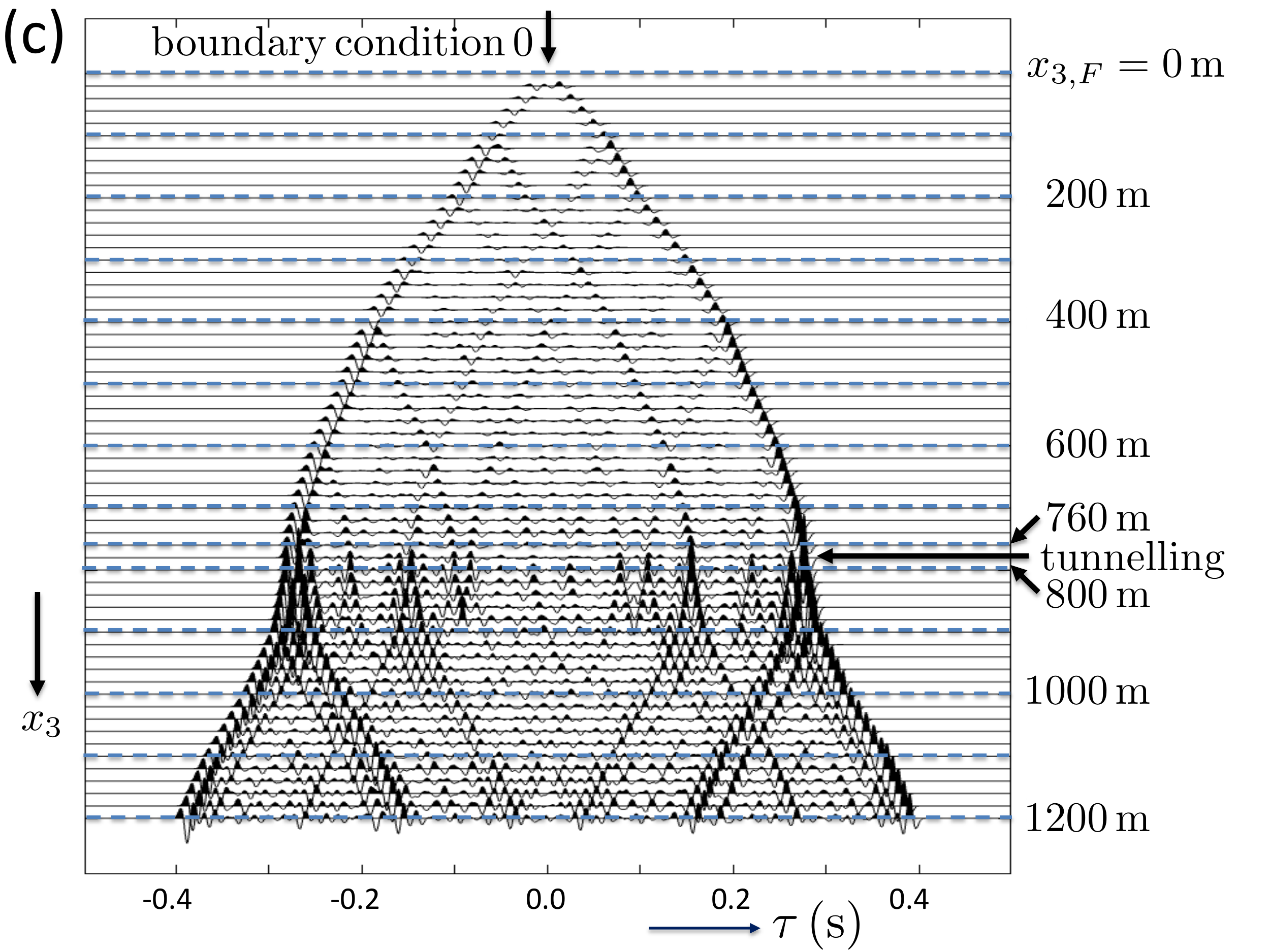}}
\vspace{0.cm}
\caption{(a) Horizontally layered medium.  (b) Propagator matrix component $W^{p,p}(s_1,x_3,x_{3,F},\tau)$ (for fixed $s_1=1/3500$ m/s).
(c) Propagator matrix component $W^{p,v}(s_1,x_3,x_{3,F},\tau)$. }\label{Figure2}
\end{figure}

We conclude this section with a numerical illustration of the propagator matrix for the horizontally layered lossless medium of Figure \ref{Figure2}(a).
In each layer the propagation velocity $c=1/\sqrt{\kappa\rho}$ is shown (in m/s). 
We define the spatial Fourier transformation of a function $u({\bf x},\omega)$ along the horizontal coordinate ${\bf x}_{\rm H}$ for constant $x_3$ as
\begin{eqnarray}
\tilde u({{\bf s}},x_3,\omega)=\int_{{\mathbb{R}}^2}\exp\{-i\omega{{\bf s}}\cdot{{\bf x}_{\rm H}}\}u({{\bf x}_{\rm H}},x_3,\omega){\rm d}^2{{\bf x}_{\rm H}},\label{eq99950b}
\end{eqnarray}
where ${{\bf s}}=(s_1,s_2)$ is the horizontal slowness vector and ${\mathbb{R}}$ is the set of real numbers. 
For a horizontally layered medium, this transformation decomposes $u({\bf x},\omega)$ into independent 
plane waves, with propagation angle $\theta$ (with respect to the vertical axis) obeying $\sin\theta=c|{\bf s}|$.
We apply this transformation to the propagator matrix ${\bf W}({\bf x},{\bf x}_F)$,  choosing ${\bf x}_F=(0,0,x_{3,F})$.
This yields the transformed propagator matrix $\tilde{\bf W}({\bf s},x_3,x_{3,F})$, with boundary condition $\tilde{\bf W}({\bf s},x_{3,F},x_{3,F})={\bf I}$. 
Analogous to equation (\ref{eq1330k}) it obeys the recursive expression
\begin{eqnarray}
\tilde{\bf W}({\bf s},x_3',x_{3,F})=\tilde{\bf W}({\bf s},x_3',x_3)\tilde{\bf W}({\bf s},x_3,x_{3,F}).\label{eqWrecom}
\end{eqnarray}
Next, we define the inverse temporal Fourier transformation for constant ${\bf s}$ and $x_3$ as
\begin{eqnarray}
u({{\bf s}},x_3,\tau)=\frac{1}{\pi}\Re\int_0^\infty \tilde u({{\bf s}},x_3,\omega)\exp\{-i\omega\tau\}{\rm d}\omega,\label{eq500a}
\end{eqnarray}
where $\Re$ denotes the real part and $\tau$ is the intercept time \citep{Stoffa89Book}. 
Applying this transformation to $\tilde{\bf W}({\bf s},x_3,x_{3,F})$ we obtain ${\bf W}({\bf s},x_3,x_{3,F},\tau)$, with boundary condition ${\bf W}({\bf s},x_{3,F},x_{3,F},\tau)={\bf I}\delta(\tau)$ 
and with ${\bf W}({\bf s},x_3,x_{3,F},\tau)$ obeying the recursive expression
\begin{eqnarray}
{\bf W}({\bf s},x_3',x_{3,F},\tau)={\bf W}({\bf s},x_3',x_3,\tau)*{\bf W}({\bf s},x_3,x_{3,F},\tau),\label{eqWrectau}
\end{eqnarray}
where the inline asterisk denotes temporal convolution.
Although the numerical modeling is most efficiently done in the slowness-frequency domain (using equation (\ref{eqWrecom})), the results are more conveniently
interpreted when displayed in the slowness intercept-time domain. Setting $s_2=0$, the
components $W^{p,p}(s_1,x_3,x_{3,F},\tau)$ and $W^{p,v}(s_1,x_3,x_{3,F},\tau)$, 
with boundary conditions $W^{p,p}(s_1,x_{3,F},x_{3,F},\tau)=\delta(\tau)$ and $W^{p,v}(s_1,x_{3,F},x_{3,F},\tau)=0$,
are shown in Figures \ref{Figure2}(b) and \ref{Figure2}(c) for fixed $s_1=1/3500$ s/m, as a function of intercept time $\tau$ and depth $x_3$. To get a smooth display,
at each depth the components are convolved with a Ricker wavelet with a central frequency of 50 Hz. The upper traces at $x_3=x_{3,F}=0$ m represent the aforementioned boundary conditions.
Note that $W^{p,p}$ and $W^{p,v}$ are, for each depth $x_3$, even and odd functions, respectively, of intercept time $\tau$.
The recursive character, described by equation (\ref{eqWrectau}), is manifest in Figures \ref{Figure2}(b) and \ref{Figure2}(c).
The propagation velocity in the layer between $x_3=760$ m and  $x_3=800$ m equals 3600 m/s, which implies that for the chosen horizontal slowness $s_1=1/3500$ s/m 
we have $\sin\theta=cs_1>1$
(i.e., $\theta$ is complex-valued). Hence, waves become `evanescent' in this layer. 
The wave field tunnels through this layer and the amplitudes below this layer are higher than above it. 
In general, evanescent field components of the  propagator matrix should be handled with care, 
because next to exponentially decaying terms they contain exponentially growing terms that may cause numerical inaccuracies
\citep{Kennett79GJRAS}. In practice this means that beyond a certain horizontal slowness the wave field should be tapered to zero.

\subsection{Relation between acoustic propagator matrix and Marchenko focusing functions}\label{sec2.3}

From here onward we assume that the medium at and above $\partial\mathbb{D}_F$ is homogeneous and may be dissipative, with mass density $\rho_0$ and propagation velocity $c_0$.
The medium below $\partial\mathbb{D}_F$ may be inhomogeneous and dissipative, and it is source-free.

In preparation for defining the focusing functions, we decompose operator matrix ${\mbox{\boldmath ${\cal A}$}}$ (in the space-frequency domain) as follows
\begin{eqnarray}
{{{\mbox{\boldmath ${\cal A}$}}}}={{{\mbox{\boldmath ${\cal L}$}}}}{\bf \Lambda}{{{\mbox{\boldmath ${\cal L}$}}}}^{-1},
\end{eqnarray}
with 
\begin{eqnarray}
&&{\bf \Lambda}=\begin{pmatrix}i{\cal H}_1&0\\0&-i{\cal H}_1\end{pmatrix},\quad {\cal H}_1=\rho^{1/2}{\cal H}_2^{1/2}\rho^{-1/2},\label{eqkk18}\\
&&{{{\mbox{\boldmath ${\cal L}$}}}}=\begin{pmatrix}1 & 1\\ \frac{1}{\omega\rho}{\cal H}_1 & -\frac{1}{\omega\rho}{\cal H}_1\end{pmatrix},\quad
{{{\mbox{\boldmath ${\cal L}$}}}}^{-1}=\frac{1}{2}\begin{pmatrix}1 & \omega{\cal H}_1^{-1}\rho \\1 &-\omega{\cal H}_1^{-1}\rho\end{pmatrix}\label{eqkk19}
\end{eqnarray}
\citep{Corones75JMAA, Fishman84JMP, Wapenaar86GP2, Hoop96JMP}.
The square-root operator ${\cal H}_2^{1/2}$ is symmetric in the following sense
\begin{eqnarray}
&&\int_{{\mathbb{R}}^2}\{{\cal H}_2^{1/2}g({\bf x}_{\rm H})\}h({\bf x}_{\rm H}){\rm d}^2{\bf x}_{\rm H}=
\int_{{\mathbb{R}}^2}g({\bf x}_{\rm H})\{{\cal H}_2^{1/2}h({\bf x}_{\rm H})\}{\rm d}^2{\bf x}_{\rm H}
\label{eq20kk}
\end{eqnarray}
\citep{Wapenaar2001RS}, where $g({\bf x}_{\rm H})$ and $h({\bf x}_{\rm H})$ are test functions in the horizontal plane with 
`sufficient decay at infinity'. Operator ${\cal H}_1$, as defined in equation (\ref{eqkk18}) is not symmetric,
but operator $\frac{1}{\rho}{\cal H}_1$ and its inverse, both appearing in equation (\ref{eqkk19}), are symmetric.
We use the operator matrix ${{{\mbox{\boldmath ${\cal L}$}}}}$ to express the wave field vector ${\bf q}({\bf x})$ in terms of downgoing and upgoing waves $p^+({\bf x})$ and $p^-({\bf x})$ via
\begin{eqnarray}\label{eq15ffrev}
{\bf q}({\bf x})={{{\mbox{\boldmath ${\cal L}$}}}}({\bf x}){\bf p}({\bf x}),
\end{eqnarray}
with
\begin{eqnarray}
{\bf p}({\bf x})=\begin{pmatrix}p^+ \\p^-\end{pmatrix}({\bf x}).\label{eq517rev}
\end{eqnarray}
Note that these equations imply 
\begin{eqnarray}\label{eqdecomfield}
p({\bf x})=p^+({\bf x}) + p^-({\bf x}),
\end{eqnarray}
hence, the downgoing and upgoing waves $p^+$ and $p^-$ are pressure-normalised.
An advantage of pressure-normalised (or, more generally, field-normalised) decomposition is that the decomposed quantities simply add up to a field quantity 
(acoustic pressure in the case of equation (\ref{eqdecomfield})).
This property does not apply to flux-normalised decomposed wave fields \citep{Frasier70GEO, Kennett78GJRAS, Ursin83GEO}.
On the other hand, an advantage of flux-normalised decomposition is that the underlying equations obey more simple symmetry properties. 
For a comprehensive discussion on field-normalised versus flux-normalised decomposition in inhomogeneous media, see \cite{Hoop96JMP} and \cite{Wapenaar2020AMP}.
In this paper we use field-normalised decomposition. 
In the remainder of section \ref{sec2} we apply decomposition only at and above $\partial\mathbb{D}_F$, where the medium is assumed to be homogeneous.
In section \ref{sec3} we will apply decomposition also inside the inhomogeneous medium.

We use equation (\ref{eq15ffrev}) at $\partial\mathbb{D}_F$ 
to derive focusing functions and express them in the components of the propagator matrix and vice-versa. Substituting 
equation (\ref{eq15ffrev}), with ${\bf x}$ replaced by ${\bf x}_F$, into the right-hand side of equation (\ref{eq1330}) gives
\begin{eqnarray}
{\bf q}({\bf x})=\int_{\partial\mathbb{D}_F} {\bf Y}({\bf x},{\bf x}_F){\bf p}({\bf x}_F){\rm d}^2{\bf x}_F,\label{eq1330dec}
\end{eqnarray}
for $x_3\ge x_{3,F}$, with 
\begin{eqnarray}
{\bf Y}({\bf x},{\bf x}_F)={\bf W}({\bf x},{\bf x}_F){{{{\mbox{\boldmath ${\cal L}$}}}}}({\bf x}_F),\label{eq15k}
\end{eqnarray}
or
\begin{eqnarray}
&&{\bf Y}({\bf x},{\bf x}_F)= \begin{pmatrix}W^{p,p} & W^{p,v} \\ W^{v,p} & W^{v,v} \end{pmatrix}({\bf x},{\bf x}_F)
\begin{pmatrix}1 & 1\\ \frac{1}{\omega\rho_0}{\cal H}_1 & -  \frac{1}{\omega\rho_0}{\cal H}_1\end{pmatrix}({\bf x}_F).\label{eq519}
\end{eqnarray}
The operators $\pm\frac{1}{\omega\rho_0}{\cal H}_1({\bf x}_F)$ in equation (\ref{eq519}) act, via equation (\ref{eq1330dec}), on $p^\pm({\bf x}_F)$.
However, since these operators are symmetric (in the sense of equation (\ref{eq20kk})), we may replace the actions of these operators on $p^\pm({\bf x}_F)$ 
by actions on the elements $W^{p,v}({\bf x},{\bf x}_F)$ and $W^{v,v}({\bf x},{\bf x}_F)$. To be more specific, if we partition ${\bf Y}({\bf x},{\bf x}_F)$ as follows
\begin{eqnarray}
{\bf Y}({\bf x},{\bf x}_F)=\begin{pmatrix}Y^{p,+} & Y^{p,-} \\ Y^{v,+}  & Y^{v,-}  \end{pmatrix}({\bf x},{\bf x}_F),\label{eq22nn}
\end{eqnarray}
we obtain from equation (\ref{eq519}) for the elements of this matrix
\begin{eqnarray}
&&Y^{p,\pm}({\bf x},{\bf x}_F)=W^{p,p}({\bf x},{\bf x}_F)\pm\frac{1}{\omega\rho_0}{\cal H}_1({\bf x}_F)W^{p,v}({\bf x},{\bf x}_F),\label{eq11}\\
&&Y^{v,\pm}({\bf x},{\bf x}_F)=W^{v,p}({\bf x},{\bf x}_F)\pm\frac{1}{\omega\rho_0}{\cal H}_1({\bf x}_F)W^{v,v}({\bf x},{\bf x}_F).\label{eq11v}
\end{eqnarray}
We analyse these expressions one by one. First we consider the element $Y^{p,-}$.
From equations (\ref{eq1330dec}) and (\ref{eq22nn}) it can be seen that the superscript $p$ refers to the acoustic pressure $p({\bf x})$ contained in ${\bf q}({\bf x})$ 
and superscript $-$ refers to the upgoing wave field component $p^-({\bf x}_F)$ in ${\bf p}({\bf x}_F)$.
Using equations (\ref{eq9998d}), (\ref{eq424}) and (\ref{eq11}) we obtain 
\begin{eqnarray}
Y^{p,-}({\bf x},{\bf x}_F)|_{x_3=x_{3,F}} = \delta({\bf x}_{\rm H}-{\bf x}_{{\rm H},F}),\label{eq9998dag}
\end{eqnarray}
which is a focusing condition.
Hence, we define 
\begin{eqnarray}
&&Y^{p,-}({\bf x},{\bf x}_F)=F^p({\bf x},{\bf x}_F)
=W^{p,p}({\bf x},{\bf x}_F)-\frac{1}{\omega\rho_0}{\cal H}_1({\bf x}_F)W^{p,v}({\bf x},{\bf x}_F),\label{eq15b}
\end{eqnarray}
with $F^p({\bf x},{\bf x}_F)$ denoting a focusing function for the acoustic pressure $p$, which focuses 
as an upgoing field at ${\bf x}={\bf x}_F$ and continues as an upgoing field in the homogeneous upper half-space,
see the lower frame of Figure \ref{Fig1}.
Next, we consider the element $Y^{v,-}$. Superscript $v$ refers to the vertical particle velocity $v_3({\bf x})$ contained in ${\bf q}({\bf x})$  and superscript
$-$ refers again to the upgoing wavefield component  $p^-({\bf x}_F)$ in ${\bf p}({\bf x}_F)$. 
Using equations (\ref{eq9998d}), (\ref{eq424}) and (\ref{eq11v}) we obtain 
\begin{eqnarray}
Y^{v,-}({\bf x},{\bf x}_F)|_{x_3=x_{3,F}} = -\frac{1}{\omega\rho_0}{\cal H}_1({\bf x}_F)\delta({\bf x}_{\rm H}-{\bf x}_{{\rm H},F}),\label{eq9998dagg}
\end{eqnarray}
which is also a focusing condition, but somewhat more complicated than equation (\ref{eq9998dag}) because of the mix of the involved wavefield components $v_3({\bf x})$ and $p^-({\bf x}_F)$. 
Hence, we define
\begin{eqnarray}
&&Y^{v,-}({\bf x},{\bf x}_F)=F^v({\bf x},{\bf x}_F)
=W^{v,p}({\bf x},{\bf x}_F)-\frac{1}{\omega\rho_0}{\cal H}_1({\bf x}_F)W^{v,v}({\bf x},{\bf x}_F),\label{eq15bb}
\end{eqnarray}
with $F^v({\bf x},{\bf x}_F)$ denoting the particle velocity counterpart of the focusing function $F^p({\bf x},{\bf x}_F)$
(note that the definition of $F^v({\bf x},{\bf x}_F)$ is different from that in \cite{Wapenaar2022JASA}, to facilitate the derivations below).
The focusing functions $F^p({\bf x},{\bf x}_F)$ and $F^v({\bf x},{\bf x}_F)$, which together form the right column of matrix ${\bf Y}({\bf x},{\bf x}_F)$, 
are illustrated in the lower frame of Figure \ref{Fig1}. They resemble the focusing function $f_2$ introduced in previous work \citep{Wapenaar2013PRL, Slob2014GEO},
which also focuses at the upper boundary (as opposed to the focusing function $f_1$, which focuses inside the medium). However, there are also some notable differences. 
First, $f_2({\bf x},{\bf x}_F)$ is defined in a truncated version of the actual medium and is obtained from a superposition of downgoing and upgoing components, $f_2^+({\bf x},{\bf x}_F)$
and $f_2^-({\bf x},{\bf x}_F)$ respectively, at ${\bf x}$ inside the medium (at the lower boundary of the truncated medium). 
Moreover, representations involving $f_2^+$ and $f_2^-$ ignore evanescent waves at $x_{3,F}$ and $x_3$. In contrast, 
$F^p({\bf x},{\bf x}_F)$ and $F^v({\bf x},{\bf x}_F)$ are defined in the actual (i.e., untruncated) 
medium and represent the full pressure and vertical particle velocity at ${\bf x}$ of a field that focuses at ${\bf x}_F$ at the upper boundary.
Since they are derived from the propagator matrix, these focusing functions account for evanescent waves
(this will be demonstrated below with a numerical example). The only decomposition takes place at the boundary $\partial\mathbb{D}_F$, where the medium is homogeneous.
This decomposition, formulated by equations (\ref{eq15b}) and (\ref{eq15bb}), accounts for evanescent waves. 
Last but not least, $F^p$ and $F^v$ hold for dissipative media and they are normalised differently from $f_2$. 

Before we analyse the elements in the left column of matrix ${\bf Y}({\bf x},{\bf x}_F)$, we introduce an adjoint medium, with parameters $\bar\kappa({\bf x})=\kappa^*({\bf x})$ and
$\bar\rho({\bf x})=\rho^*({\bf x})$. The bar denotes the adjoint medium and the superscript 
asterisk denotes complex conjugation. When the original medium is dissipative, the adjoint medium
is effectual, with (for positive $\omega$) $\Im(\bar\kappa)\le 0$ and $\Im(\bar\rho)\le 0$. Waves propagating through an effectual medium gain energy \citep{Bojarski83JASA, Hoop88JASA}.
Adjoint media are usually associated to a computational state. The operator matrix 
${\,\,\,\bar{\mbox{\!\!\!\boldmath ${\cal A}$}}}$ and the Helmholtz operator $\bar{\cal H}_2$ of the adjoint medium are defined similarly as ${{\mbox{\boldmath ${\cal A}$}}}$ and ${\cal H}_2$
in equations (\ref{eqAcoustic}) and (\ref{eqHelmholtz}), respectively, but with $\kappa({\bf x})$ and $\rho({\bf x})$ replaced by $\bar\kappa({\bf x})$ and $\bar\rho({\bf x})$, respectively.
Hence, $\bar{\cal H}_2={\cal H}_2^*$.
Analogous to equations (\ref{eq2.1}) and (\ref{eq9998d}), we define the propagator matrix $\bar{\bf W}({\bf x},{\bf x}_F)$ of the adjoint medium as the solution of 
$\partial_3\bar{\bf W}({\bf x},{\bf x}_F) = {\,\,\,\bar{\mbox{\!\!\!\boldmath ${\cal A}$}}}({\bf x})\bar{\bf W}({\bf x},{\bf x}_F)$, with boundary condition
$\bar{\bf W}({\bf x},{\bf x}_F)|_{x_3={x_{3,F}}} = {\bf I}\delta({{\bf x}_{\rm H}}-{{\bf x}_{{\rm H},F}})$.
In Appendix \ref{AppC} we derive
\begin{eqnarray}
\begin{pmatrix}\bar W^{p,p}      & \bar W^{p,v} \\
                \bar W^{v,p} &  \bar W^{v,v}    \end{pmatrix}({\bf x},{\bf x}_F)=
                \begin{pmatrix}W^{p,p*}      & -W^{p,v*} \\
                -W^{v,p*} & W^{v,v*}    \end{pmatrix}({\bf x},{\bf x}_F)\label{eq7}
\end{eqnarray}
(equation (\ref{eq65aws})).
For the square-root operator we have, similar as for the Helmholtz operator,
\begin{eqnarray}
\bar{\cal H}_1={\cal H}_1^*\label{eq9}
\end{eqnarray}
\citep{Wapenaar2001RS}. Using equations (\ref{eq7}) and (\ref{eq9}) in equations (\ref{eq11}) and (\ref{eq11v}), we find $\bar Y^{p,+}({\bf x},{\bf x}_F)=Y^{p,-*}({\bf x},{\bf x}_F)$ 
and $\bar Y^{v,+}({\bf x},{\bf x}_F)=-Y^{v,-*}({\bf x},{\bf x}_F)$. Hence, using equations (\ref{eq15b}) and (\ref{eq15bb}), we find for the elements in the left column of matrix ${\bf Y}({\bf x},{\bf x}_F)$
\begin{eqnarray}
&&Y^{p,+}({\bf x},{\bf x}_F)=\bar F^{p*}({\bf x},{\bf x}_F)
=W^{p,p}({\bf x},{\bf x}_F)+\frac{1}{\omega\rho_0}{\cal H}_1({\bf x}_F)W^{p,v}({\bf x},{\bf x}_F),\label{eq16}\\
&&Y^{v,+}({\bf x},{\bf x}_F)=-\bar F^{v*}({\bf x},{\bf x}_F)
=W^{v,p}({\bf x},{\bf x}_F)+\frac{1}{\omega\rho_0}{\cal H}_1({\bf x}_F)W^{v,v}({\bf x},{\bf x}_F).\label{eq16bb}
\end{eqnarray}
For matrix ${\bf Y}({\bf x},{\bf x}_F)$ we thus obtain
\begin{eqnarray}
{\bf Y}({\bf x},{\bf x}_F)= \begin{pmatrix}\bar F^{p*} & F^p \\ -\bar F^{v*} & F^v \end{pmatrix}({\bf x},{\bf x}_F).\label{eq21}
\end{eqnarray}
Note that $F^p$, $F^v$, $\bar F^{p*}$ and $\bar F^{v*}$ are expressed in terms of the components of the propagator matrix ${\bf W}({\bf x},{\bf x}_F)$ via equations
(\ref{eq15b}), (\ref{eq15bb}), (\ref{eq16}) and (\ref{eq16bb}). Conversely,
we can express the components of the propagator matrix ${\bf W}({\bf x},{\bf x}_F)$ in terms of the focusing functions $F^p$, $F^v$, $\bar F^{p*}$ and $\bar F^{v*}$.
Inverting equation (\ref{eq15k}) yields
\begin{eqnarray}
{\bf W}({\bf x},{\bf x}_F)={\bf Y}({\bf x},{\bf x}_F){{{{\mbox{\boldmath ${\cal L}$}}}}}^{-1}({\bf x}_F),\label{eqWYD}
\end{eqnarray}
with ${{{{\mbox{\boldmath ${\cal L}$}}}}}^{-1}$ defined in equation (\ref{eqkk19}).
Since operator $\frac{1}{\rho}{\cal H}_1$ is symmetric, its inverse ${\cal H}_1^{-1}\rho$ is symmetric as well. 
Hence, in equation (\ref{eqWYD}) these operators can be taken to act on the elements of matrix ${\bf Y}({\bf x},{\bf x}_F)$. This yields 
\begin{eqnarray}
&&W^{p,p}({\bf x},{\bf x}_F) = \frac{1}{2}\bigl( \bar F^{p*} + F^p\bigr)({\bf x},{\bf x}_F),\label{eq14}\\
&&W^{p,v}({\bf x},{\bf x}_F)=\frac{\omega\rho_0}{2}{\cal H}_1^{-1}({\bf x}_F)\bigl( \bar F^{p*} - F^p\bigr)({\bf x},{\bf x}_F),\label{eq15}\\
&&W^{v,p}({\bf x},{\bf x}_F) = \frac{1}{2}\bigl( -\bar F^{v*} + F^v\bigr)({\bf x},{\bf x}_F),\label{eq14v}\\
&&W^{v,v}({\bf x},{\bf x}_F)=-\frac{\omega\rho_0}{2}{\cal H}_1^{-1}({\bf x}_F)\bigl( \bar F^{v*} + F^v\bigr)({\bf x},{\bf x}_F).\label{eq15v}
\end{eqnarray}
Note that up to this point the medium may be dissipative (and its adjoint effectual), 
and evanescent wave modes are accounted for, inside the medium as well as at the boundary $\partial\mathbb{D}_F$.
Hence, the expressions in this section are more general than their counterparts in \cite{Wapenaar2022JASA}, which were derived for a lossless medium, 
under the assumption that evanescent waves can be ignored at $\partial\mathbb{D}_F$. If we make the same assumptions here, we can omit the bars on $F^p$ and $F^v$.
For this situation equations (\ref{eq14}) -- (\ref{eq15v}) simplify to
\begin{eqnarray}
&&W^{p,p}({\bf x},{\bf x}_F) = \Re \{F^p({\bf x},{\bf x}_F)\},\label{eq14g}\\
&&W^{p,v}({\bf x},{\bf x}_F)=-i\omega\rho_0{\cal H}_1^{-1}({\bf x}_F)\Im\{F^p({\bf x},{\bf x}_F)\},\label{eq15g}\\
&&W^{v,p}({\bf x},{\bf x}_F) = i\Im\{F^v\bigr({\bf x},{\bf x}_F)\},\label{eq14vg}\\
&&W^{v,v}({\bf x},{\bf x}_F)=-\omega\rho_0{\cal H}_1^{-1}({\bf x}_F)\Re\{F^v({\bf x},{\bf x}_F)\}.\label{eq15vg}
\end{eqnarray}

\begin{figure}
\centerline{\epsfysize=7.cm \epsfbox{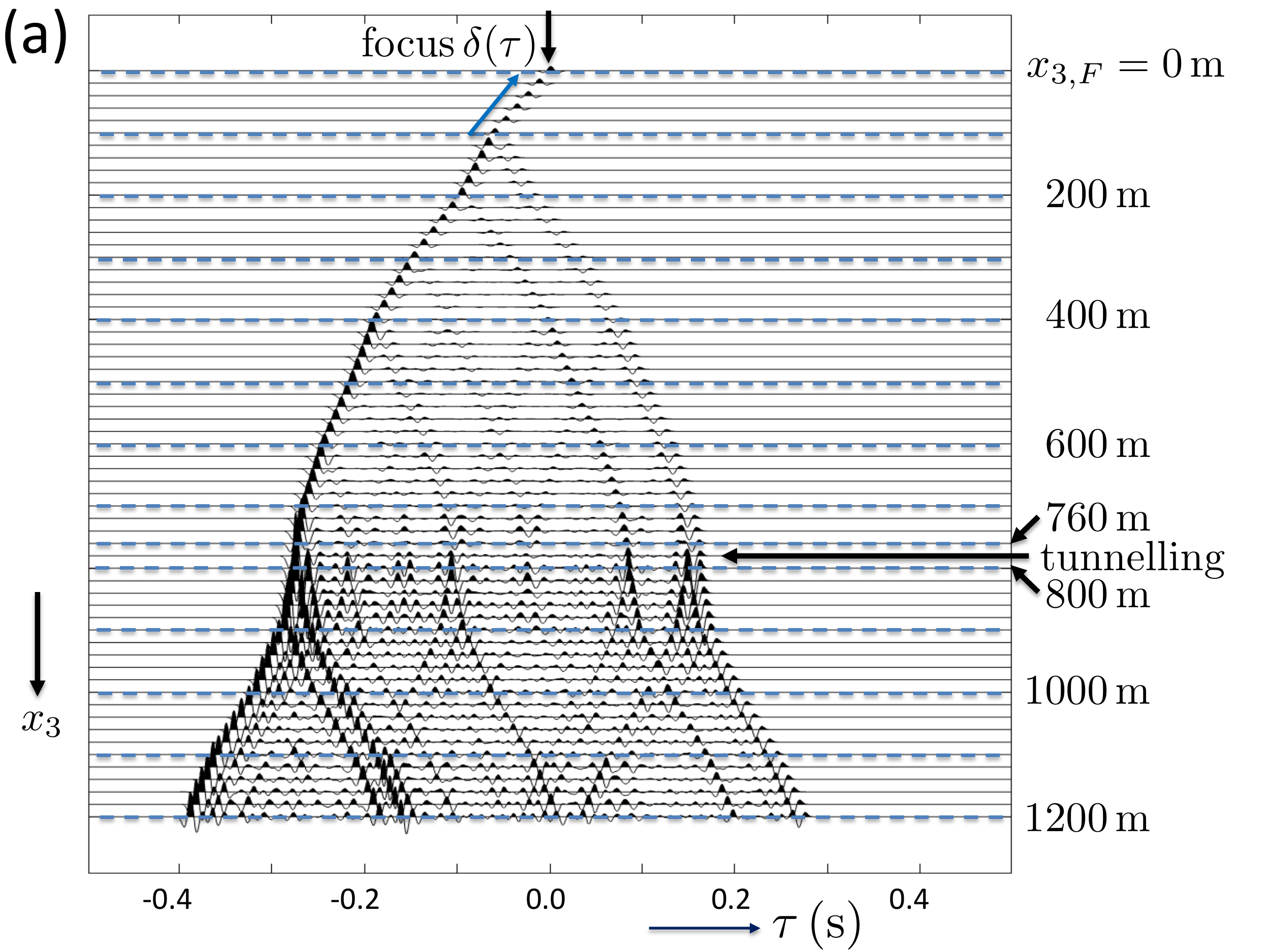}}
\vspace{0.4cm}
\centerline{\epsfysize=7. cm \epsfbox{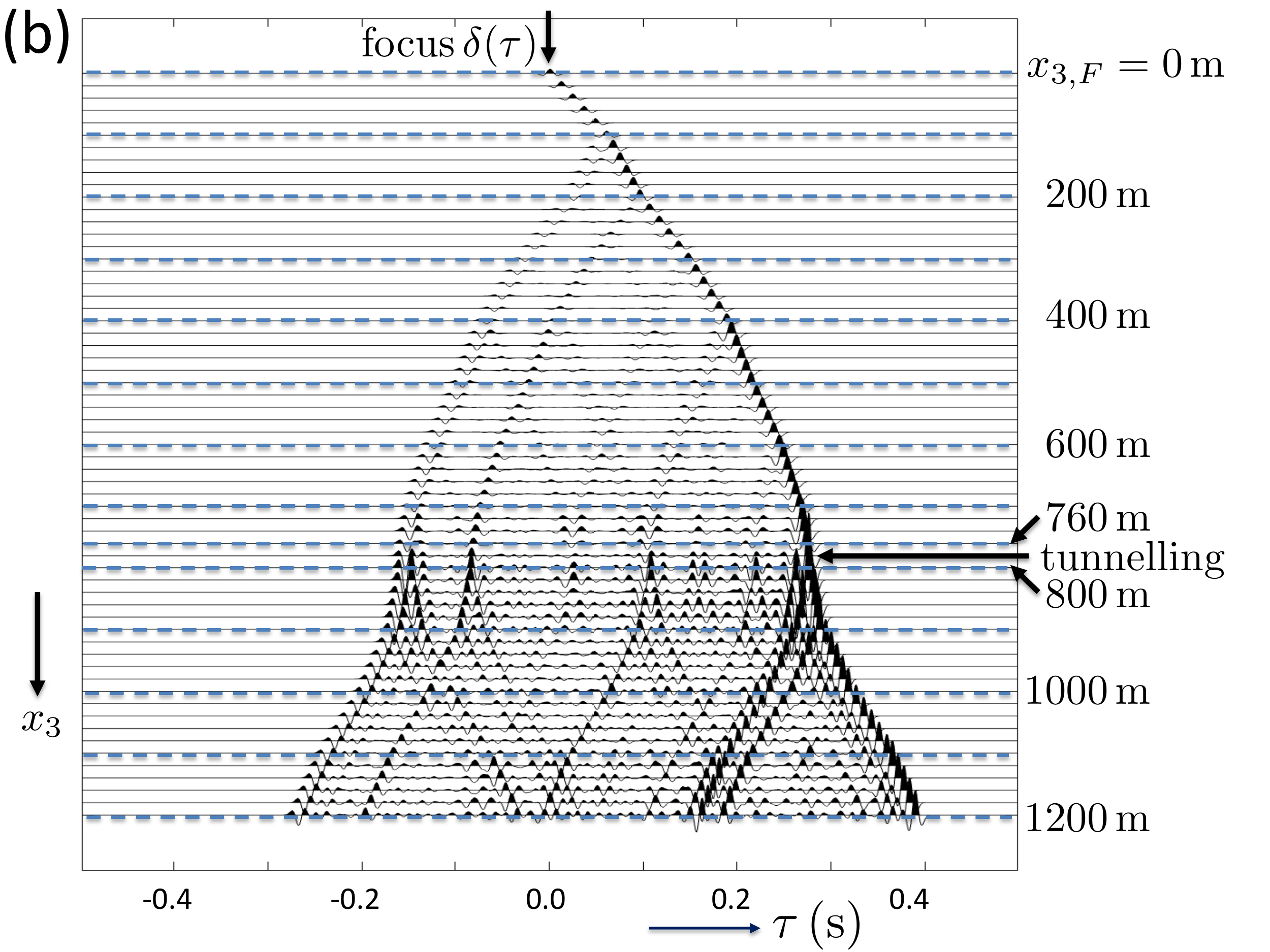}}
\vspace{0.cm}
\caption{(a) Focusing function $F^p(s_1,x_3,x_{3,F},\tau)$ (for fixed $s_1=1/3500$ m/s).
(b) Time-reversed focusing function $F^p(s_1,x_3,x_{3,F},-\tau)$. }\label{Figure4}
\end{figure}

We illustrate the focusing function and its relation with the propagator matrix with a numerical example.
Applying the transformations of equations (\ref{eq99950b}) and (\ref{eq500a}) to equation (\ref{eq15b}) (assuming a laterally invariant medium), taking 
 ${\bf x}_F=(0,0,x_{3,F})$ and  $s_2=0$, we obtain 
\begin{eqnarray}\label{eq41W}
F^p(s_1,x_3,x_{3,F},\tau)&=&W^{p,p}(s_1,x_3,x_{3,F},\tau)
-\frac{s_{3,0}}{\rho_0}W^{p,v}(s_1,x_3,x_{3,F},\tau),
\end{eqnarray}
with vertical slowness $s_{3,0}=\sqrt{1/c_0^2-s_1^2}$ being the spatial Fourier transform of $\frac{1}{\omega}{\cal H}_1$ at $x_{3,F}$ for the laterally invariant medium
(here we assumed $s_1^2<1/c_0^2$).
Equation (\ref{eq41W}) shows how a weighted superposition of the even component $W^{p,p}$ of Figure \ref{Figure2}(b) and the odd component $W^{p,v}$ of Figure \ref{Figure2}(c)
yields the focusing function $F^p(s_1,x_3,x_{3,F},\tau)$. This focusing function is shown in Figure \ref{Figure4}(a) for $s_1=1/3500$ m/s. 
The upper trace at $x_3=x_{3,F}=0$ m represents the focusing condition  $F^p(s_1,x_{3,F},x_{3,F},\tau)=\delta(\tau)$. 
At and above $x_{3,F}$ the focusing function is an upgoing field.
Note that, similar as in Figure \ref{Figure2}, the wave field tunnels through the high-velocity layer between $x_3=760$ m and $x_3=800$ m, 
which confirms that this focusing function accounts for evanescent waves inside the medium.
The time-reversed focusing function $F^p(s_1,x_3,x_{3,F},-\tau)$ is shown in Figure \ref{Figure4}(b). 
The focusing function of Figure \ref{Figure4}(a) and its time-reversed version of Figure \ref{Figure4}(b) can be combined to give components of the propagator matrix.
To this end, equations (\ref{eq14g}) and (\ref{eq15g}) are transformed to (assuming $s_1^2<1/c_0^2$) 
 \begin{eqnarray}\label{eq42W}
 &&W^{p,p}(s_1,x_3,x_{3,F},\tau)=
 \frac{1}{2}\Bigl(F^p(-s_1,x_3,x_{3,F},-\tau)+F^p(s_1,x_3,x_{3,F},\tau)\Bigr),\label{eq43FW}\\
&&W^{p,v}(s_1,x_3,x_{3,F},\tau)=
\frac{\rho_0}{2s_{3,0}}\Bigl(F^p(-s_1,x_3,x_{3,F},-\tau)-F^p(s_1,x_3,x_{3,F},\tau)\Bigr).\label{eq44FW}
 \end{eqnarray}
For the acoustic case all components are symmetric in $s_1$, i.e., 
$F^p(-s_1,x_3,x_{3,F},-\tau)=F^p(s_1,x_3,x_{3,F},-\tau)$, etc.
Hence, equations (\ref{eq43FW}) and (\ref{eq44FW}) 
show how the even and odd components $W^{p,p}(s_1,x_3,x_{3,F},\tau)$ and $W^{p,v}(s_1,x_3,x_{3,F},\tau)$ of Figures \ref{Figure2}(b) and \ref{Figure2}(c)
are obtained from the focusing function $F^p(s_1,x_3,x_{3,F},\tau)$ and its time-reversal $F^p(s_1,x_3,x_{3,F},-\tau)$ of Figure \ref{Figure4}.

\subsection{Representations with acoustic Marchenko focusing functions}\label{sec2.4}

Substituting the expressions for ${\bf q}({\bf x})$, ${\bf p}({\bf x}_F)$ and ${\bf Y}({\bf x},{\bf x}_F)$ 
(equations (\ref{eq9996ge}), (\ref{eq517rev}) and (\ref{eq21})) into equation (\ref{eq1330dec}), gives
the following representations for the acoustic pressure $p({\bf x})$ and the vertical particle velocity $v_3({\bf x})$ inside the inhomogeneous medium 
\begin{eqnarray}
p({\bf x})&=&\int_{\partial\mathbb{D}_F} \bar F^{p*}({\bf x},{\bf x}_F)p^+({\bf x}_F){\rm d}^2{\bf x}_F 
+\int_{\partial\mathbb{D}_F} F^p({\bf x},{\bf x}_F)p^-({\bf x}_F){\rm d}^2{\bf x}_F,\label{eq13}\\
v_3({\bf x})&=&-\int_{\partial\mathbb{D}_F} \bar F^{v*}({\bf x},{\bf x}_F)p^+({\bf x}_F){\rm d}^2{\bf x}_F 
+\int_{\partial\mathbb{D}_F} F^v({\bf x},{\bf x}_F)p^-({\bf x}_F){\rm d}^2{\bf x}_F,\label{eq13bb}
\end{eqnarray}
for $x_3\ge x_{3,F}$. These expressions are exact and hold for dissipative media.
Equation (\ref{eq13}) is a generalisation of equation (17) of \cite{Wapenaar2022GEO} for dissipative media. 

We use equations (\ref{eq13}) and (\ref{eq13bb}) to derive representations for Green's functions between the boundary $\partial\mathbb{D}_F$ and any position ${\bf x}$ inside the medium.
To this end, we define a unit point source of vertical force at ${\bf x}_S$ just above $\partial\mathbb{D}_F$.
For the downgoing field at $\partial\mathbb{D}_F$ (i.e., just below the source), we then have $p^+({\bf x}_F)=\frac{1}{2}\delta({\bf x}_{{\rm H},F}-{\bf x}_{{\rm H},S})$, where  
${\bf x}_{{\rm H},S}$ denotes the horizontal coordinates of ${\bf x}_S$. The upgoing field at $\partial\mathbb{D}_F$ is the reflection response to this downgoing source field, hence
$p^-({\bf x}_F)=\frac{1}{2} R({\bf x}_F,{\bf x}_S)$. The field at ${\bf x}$ inside the medium is the Green's response to the source at ${\bf x}_S$, hence
 $p({\bf x})=G^{p,f}({\bf x},{\bf x}_S)$ and  $v_3({\bf x})=G^{v,f}({\bf x},{\bf x}_S)$. 
 Here the second superscript ($f$) refers to the vertical force source at ${\bf x}_S$, whereas the
 first superscripts ($p$ and $v$) refer to the observed quantities (pressure and vertical particle velocity) at  ${\bf x}$.
 Substitution of these expressions for $p^\pm({\bf x}_F)$, $p({\bf x})$ and $v_3({\bf x})$ into equations (\ref{eq13}) and (\ref{eq13bb}) gives
\begin{eqnarray}
2G^{p,f}({\bf x},{\bf x}_S)&=& \int_{\partial\mathbb{D}_F} F^p({\bf x},{\bf x}_F)R({\bf x}_F,{\bf x}_S){\rm d}^2{\bf x}_F 
+\bar F^{p*}({\bf x},{\bf x}_S),\label{eq13G}\\
2G^{v,f}({\bf x},{\bf x}_S)&=& \int_{\partial\mathbb{D}_F} F^v({\bf x},{\bf x}_F)R({\bf x}_F,{\bf x}_S){\rm d}^2{\bf x}_F 
-\bar F^{v*}({\bf x},{\bf x}_S),\label{eq13Gbb}
\end{eqnarray}
for $x_3\ge x_{3,F}$. \cite{Slob2016PRL} derived similar representations for decomposed wave fields in dissipative media. In the present derivation we only used decomposition at the  
boundary $\partial\mathbb{D}_F$ (similar as \cite{Diekmann2021PRR, Diekmann2023JASA} and \cite{Wapenaar2021GJI}). 
This implies that inside the medium the wavefield does not need to be decomposed into downgoing and upgoing waves and that evanescent waves can be present.
 
When the medium is lossless and evanescent waves are neglected at $\partial\mathbb{D}_F$, the bars on $F^p$ and $F^v$ in representations (\ref{eq13G}) and (\ref{eq13Gbb}) can be omitted.
Using the Marchenko method, these focusing functions can then be retrieved from the reflection response $R({\bf x}_F,{\bf x}_S)$ and 
a smooth background model \citep{Wapenaar2013PRL, Elison2020GJI}. 
Since representations (\ref{eq13G}) and (\ref{eq13Gbb}) account for evanescent waves inside the medium, 
the retrieved focusing functions potentially also account for evanescent waves inside the 
medium (this is subject of current research  \citep{Brackenhoff2023arXiv}). Once the focusing functions are found, they can be used to retrieve the Green's functions 
$G^{p,f}({\bf x},{\bf x}_S)$ and $G^{v,f}({\bf x},{\bf x}_S)$ 
(from equations (\ref{eq13G}) and (\ref{eq13Gbb})) 
and all components of the propagator matrix ${\bf W}({\bf x},{\bf x}_F)$ (from equations (\ref{eq14g}) -- (\ref{eq15vg})).

\section{Acoustic transfer matrix and decomposed focusing functions}\label{sec3}

\subsection{Acoustic transfer matrix}

We introduce the transfer matrix as follows.
Given the downgoing and upgoing fields $p^+({\bf x}_F)$ and $p^-({\bf x}_F)$ at the boundary $\partial\mathbb{D}_F$, 
we `transfer' these fields to downgoing and upgoing fields $p^+({\bf x})$ and $p^-({\bf x})$ at any depth level $x_3$ inside the medium using the following expression
\begin{eqnarray}
{\bf p}({\bf x})=\int_{\partial\mathbb{D}_F} {{{\mbox{\boldmath ${\cal T}$}}}}({\bf x},{\bf x}_F){\bf p}({\bf x}_F){\rm d}^2{\bf x}_F,\label{eq1330trans}
\end{eqnarray}
for $x_3\ge x_{3,F}$. Vectors ${\bf p}({\bf x}_F)$ and ${\bf p}({\bf x})$ 
contain the downgoing and upgoing fields at depth levels $x_{3,F}$ and $x_3$ (equation (\ref{eq517rev})). We call ${{{\mbox{\boldmath ${\cal T}$}}}}({\bf x},{\bf x}_F)$
 the transfer matrix, which we partition as follows
\begin{eqnarray}\label{eq424T}
{{{\mbox{\boldmath ${\cal T}$}}}}({\bf x},{\bf x}_F)= \begin{pmatrix}{\cal T}^{+,+}      & {\cal T}^{+,-} \\
                {\cal T}^{-,+} & {\cal T}^{-,-}    \end{pmatrix}({\bf x},{\bf x}_F).
\end{eqnarray}
For each component of this matrix, the superscripts refer to the propagation direction at ${\bf x}$ and at ${\bf x}_F$, respectively.
Equation (\ref{eq1330trans}) is illustrated in the upper-right frame of Figure \ref{Fig1}.

For horizontally layered media, the transfer matrix is usually built up recursively from interface to interface 
\citep{Born65Book, Katsidis2002AO, Elison2020PHD, Dukalski2022EAGE, Dukalski2022IMAGE}. Here we follow a different approach to derive an expression for 
${{{\mbox{\boldmath ${\cal T}$}}}}({\bf x},{\bf x}_F)$ for laterally varying media.
Substituting equation (\ref{eq15ffrev}) into equation (\ref{eq1330}) we obtain equation (\ref{eq1330trans}), with
\begin{eqnarray}
{{{\mbox{\boldmath ${\cal T}$}}}}({\bf x},{\bf x}_F)={{{\mbox{\boldmath ${\cal L}$}}}}^{-1}({\bf x}){\bf W}({\bf x},{\bf x}_F){{{{\mbox{\boldmath ${\cal L}$}}}}}({\bf x}_F),\label{eq31}
\end{eqnarray}
with ${{{{\mbox{\boldmath ${\cal L}$}}}}}({\bf x})$  and its inverse defined in equation (\ref{eqkk19}).
Equation (\ref{eq31}), which relates the transfer matrix ${{{\mbox{\boldmath ${\cal T}$}}}}({\bf x},{\bf x}_F)$ to the propagator matrix ${\bf W}({\bf x},{\bf x}_F)$,
is illustrated in the upper half of Figure \ref{Fig1}.
In the next section we show that the transfer matrix can be expressed in terms of decomposed Marchenko focusing functions.

\subsection{Relation between acoustic transfer matrix and decomposed Marchenko focusing functions}

From equations (\ref{eq15k}) and (\ref{eq31}) we find
\begin{eqnarray}
{{{\mbox{\boldmath ${\cal T}$}}}}({\bf x},{\bf x}_F)={{{\mbox{\boldmath ${\cal L}$}}}}^{-1}({\bf x}){\bf Y}({\bf x},{\bf x}_F).\label{eq31k}
\end{eqnarray}
According to equation (\ref{eq21}), the right column of ${\bf Y}({\bf x},{\bf x}_F)$ contains $F^p({\bf x},{\bf x}_F)$ and $F^v({\bf x},{\bf x}_F)$, 
i.e., the pressure and vertical particle velocity components at ${\bf x}$ of the focusing function.
Hence, analogous to ${\bf p}({\bf x})={{{\mbox{\boldmath ${\cal L}$}}}}^{-1}({\bf x}){\bf q}({\bf x})$, 
we obtain for the right column of ${{{\mbox{\boldmath ${\cal T}$}}}}({\bf x},{\bf x}_F)$
\begin{eqnarray}\label{eq55b}
 \begin{pmatrix} F^+({\bf x},{\bf x}_F) \\  F^-({\bf x},{\bf x}_F) \end{pmatrix}=
 \frac{1}{2}\begin{pmatrix} 1 & \omega{\cal H}_1^{-1}({\bf x})\rho({\bf x})\\1 & -\omega{\cal H}_1^{-1}({\bf x})\rho({\bf x})\end{pmatrix}
 \begin{pmatrix} F^p({\bf x},{\bf x}_F) \\  F^v({\bf x},{\bf x}_F) \end{pmatrix},
\end{eqnarray}
with $F^+({\bf x},{\bf x}_F)$ and $F^-({\bf x},{\bf x}_F)$ 
being the downgoing and upgoing parts at ${\bf x}$ of the focusing function $F^p({\bf x},{\bf x}_F)$.

According to equation (\ref{eq21}), the left column of ${\bf Y}({\bf x},{\bf x}_F)$ contains $\bar F^{p*}({\bf x},{\bf x}_F)$ and $-\bar F^{v*}({\bf x},{\bf x}_F)$.
Hence, for the left column of ${{{\mbox{\boldmath ${\cal T}$}}}}({\bf x},{\bf x}_F)$ we obtain
\begin{eqnarray}
 \frac{1}{2}\begin{pmatrix} 1 & \omega{\cal H}_1^{-1}({\bf x})\rho({\bf x})\\1 & -\omega{\cal H}_1^{-1}({\bf x})\rho({\bf x})\end{pmatrix}
 \begin{pmatrix} \bar F^{p*}({\bf x},{\bf x}_F) \\  -\bar F^{v*}({\bf x},{\bf x}_F) \end{pmatrix},
\end{eqnarray}
or, using ${\cal H}_1=\bar{\cal H}_1^*$ (equation (\ref{eq9})) and $\rho=\bar\rho^*$,
\begin{eqnarray}
 \frac{1}{2}\begin{pmatrix} 1 & -\omega\bar{\cal H}_1^{-1}({\bf x})\bar\rho({\bf x})\\1 & \omega\bar{\cal H}_1^{-1}({\bf x})\bar\rho({\bf x})\end{pmatrix}^*
 \begin{pmatrix} \bar F^p({\bf x},{\bf x}_F) \\  \bar F^v({\bf x},{\bf x}_F) \end{pmatrix}^*.
\end{eqnarray}
Comparing this with equation (\ref{eq55b}) we find that this gives a vector with $\bar F^{-*}({\bf x},{\bf x}_F)$ and $\bar F^{+*}({\bf x},{\bf x}_F)$.
This is the left column of ${{{\mbox{\boldmath ${\cal T}$}}}}({\bf x},{\bf x}_F)$. Hence, we have obtained
\begin{eqnarray}\label{eq58vv}
{{{\mbox{\boldmath ${\cal T}$}}}}({\bf x},{\bf x}_F)=\begin{pmatrix} \bar F^{-*}({\bf x},{\bf x}_F) & F^+({\bf x},{\bf x}_F)\\
 \bar F^{+*}({\bf x},{\bf x}_F) & F^-({\bf x},{\bf x}_F)\end{pmatrix},
\end{eqnarray}
see the upper-right frame of Figure \ref{Fig1}.
Hence, the transfer matrix for an inhomogeneous dissipative acoustic medium is expressed in terms of decomposed focusing functions of the medium and its adjoint. 

We consider the special case of a horizontally layered medium.
Applying the transformations of equations (\ref{eq99950b}) and (\ref{eq500a}) to equation (\ref{eq58vv}), taking  ${\bf x}_F=(0,0,x_{3,F})$, we obtain 
\begin{eqnarray}\label{eq59ff}
&&{{{\mbox{\boldmath ${\cal T}$}}}}({\bf s},x_3,x_{3,F},\tau)=
\begin{pmatrix} \bar F^-(-{\bf s},x_3,x_{3,F},-\tau) &  F^+({\bf s},x_3,x_{3,F},\tau)\\
\bar F^+(-{\bf s},x_3,x_{3,F},-\tau) & F^-({\bf s},x_3,x_{3,F},\tau)\end{pmatrix}.
\end{eqnarray}
\cite{Dukalski2022EAGE, Dukalski2022IMAGE} used a recursive approach and obtained an expression similar to equation (\ref{eq59ff}). In their derivation 
they used a path-reversal operator ${\cal P}$, which is equivalent with
(i) taking the adjoint medium, 
(ii) taking the complex conjugate (or  in the time domain taking the time-reversal) and (iii) changing the sign of the horizontal slowness. 
Hence, ${\cal P}\{F^\pm({\bf s},x_3,x_{3,F},\tau)\}$ is equivalent with $\bar F^\pm(-{\bf s},x_3,x_{3,F},-\tau)$.

\begin{figure}
\centerline{\epsfysize=7.cm \epsfbox{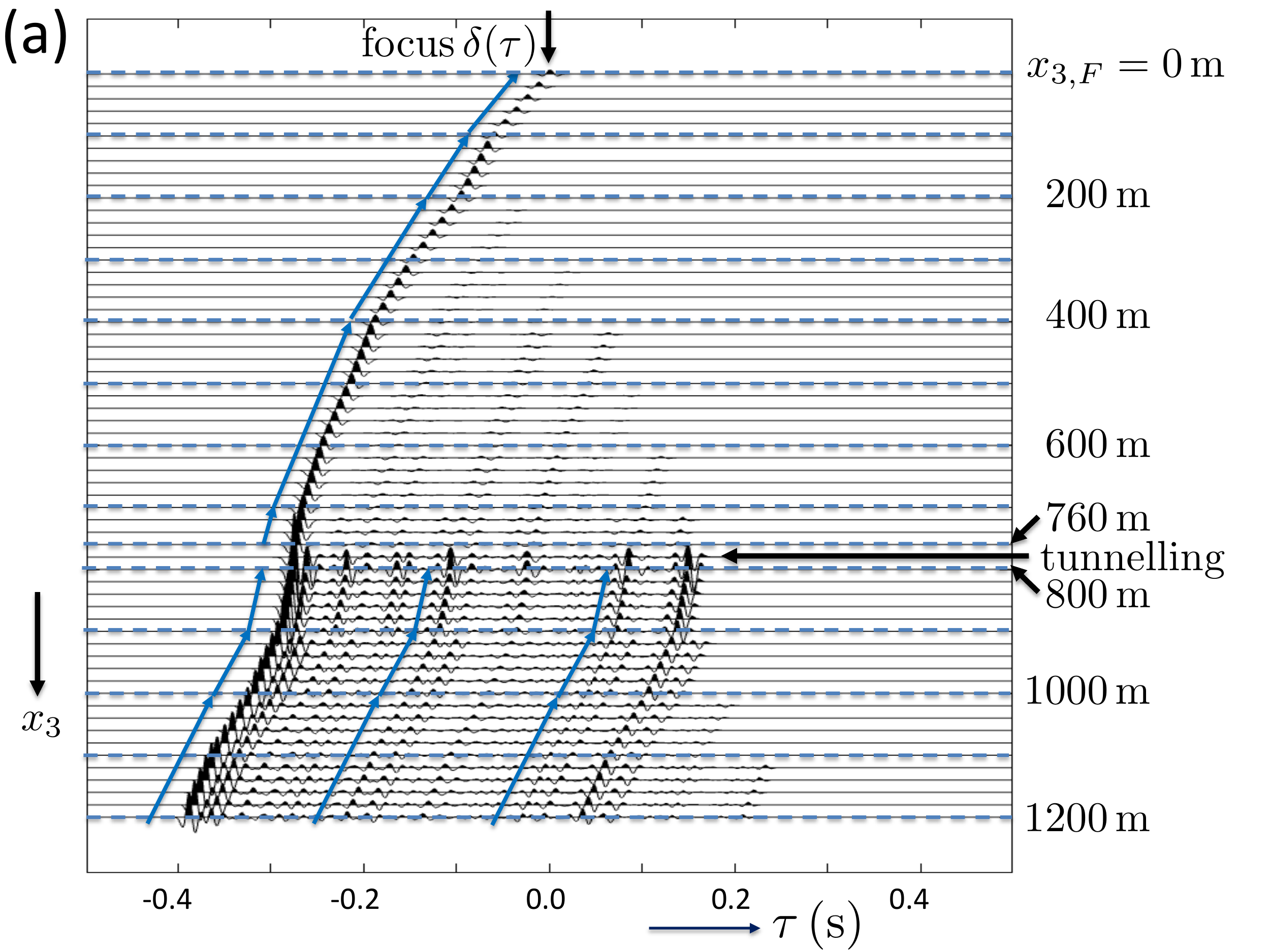}}
\vspace{.4cm}
\centerline{\epsfysize=7. cm \epsfbox{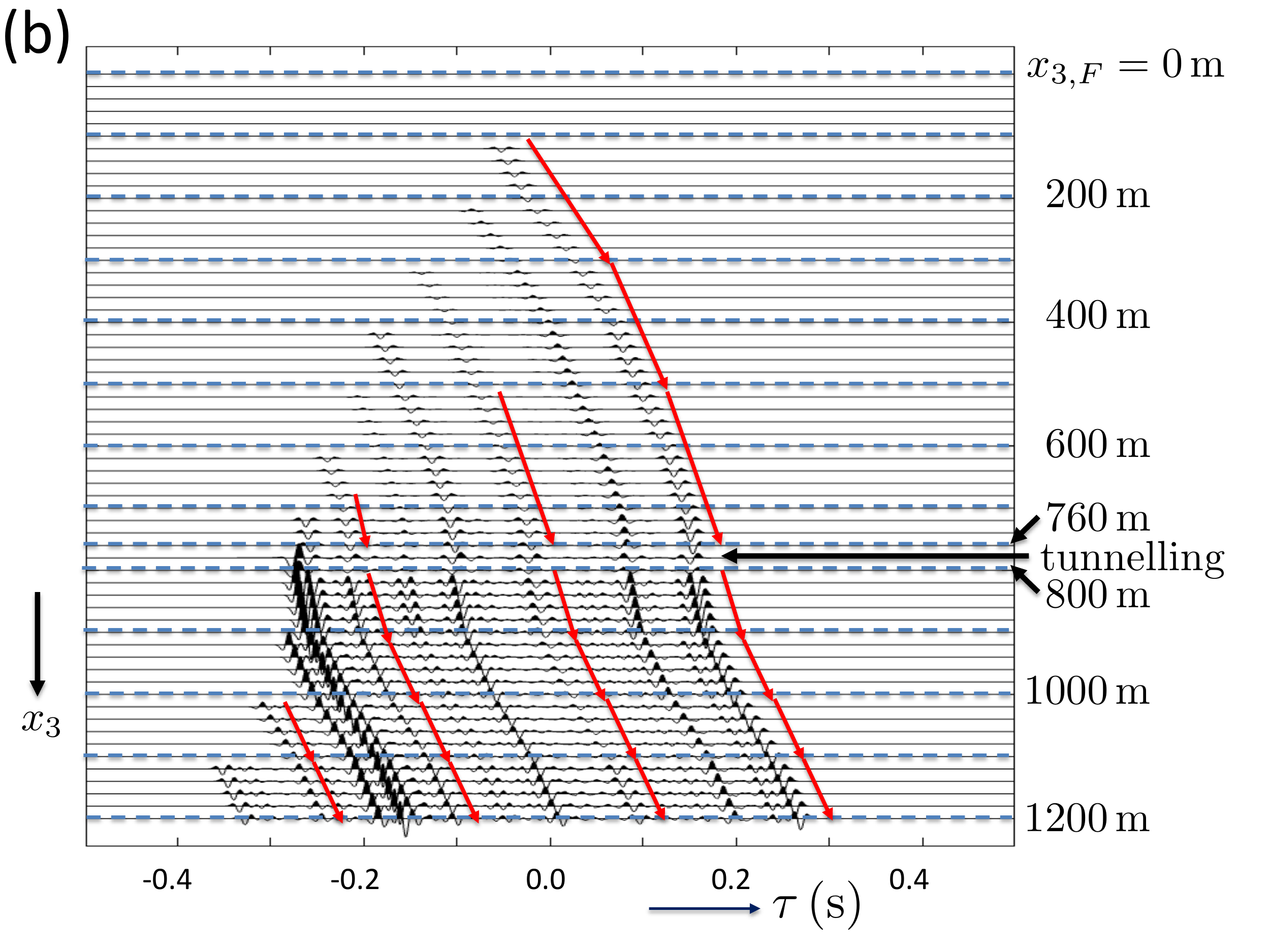}}
\vspace{0.cm}
\caption{ (a) Decomposed focusing function $F^-(s_1,x_3,x_{3,F},\tau)$ (for fixed $s_1=1/3500$ m/s).
(b) Decomposed focusing function $F^+(s_1,x_3,x_{3,F},\tau)$ . }\label{Figure5}
\end{figure}

For the lossless medium of Figure \ref{Figure2}(a), the decomposed focusing functions $F^-(s_1,x_3,x_{3,F},\tau)$ and $F^+(s_1,x_3,x_{3,F},\tau)$
for $s_1=1/3500$ m/s and $s_2=0$
are shown in Figures \ref{Figure5}(a) and \ref{Figure5}(b), respectively. For each $x_3$, the function $F^-(s_1,x_3,x_{3,F},\tau)$ can be seen as the intricate
field that needs to be emitted upward from $x_3$ to arrive as a single upward propagating field at the focal depth $x_{3,F}$ at $\tau=0$. For the same $x_3$, 
the function $F^+(s_1,x_3,x_{3,F},\tau)$ is the downward reflected response to $F^-(s_1,x_3,x_{3,F},\tau)$.
Figures \ref{Figure5}(b) and \ref{Figure5}(a) together form the right column of the transformed transfer matrix ${{{\mbox{\boldmath ${\cal T}$}}}}(s_1,x_3,x_{3,F},\tau)$.
Their superposition gives the focusing function $F^p(s_1,x_3,x_{3,F},\tau)$, shown in Figure \ref{Figure4}(a). 

\subsection{Representations with decomposed acoustic Marchenko focusing functions}\label{sec3.3}

Substituting the expressions for ${\bf p}({\bf x})$ and ${{{\mbox{\boldmath ${\cal T}$}}}}({\bf x},{\bf x}_F)$ (equations (\ref{eq517rev}) and (\ref{eq58vv})) 
into equation (\ref{eq1330trans}), gives
the following representations for the downgoing and upgoing components of the acoustic pressure, $p^+({\bf x})$ and $p^-({\bf x})$ respectively, inside the inhomogeneous medium 
\begin{eqnarray}
p^+({\bf x})&=&\int_{\partial\mathbb{D}_F} \bar F^{-*}({\bf x},{\bf x}_F)p^+({\bf x}_F){\rm d}^2{\bf x}_F 
+\int_{\partial\mathbb{D}_F} F^+({\bf x},{\bf x}_F)p^-({\bf x}_F){\rm d}^2{\bf x}_F,\label{eq13deco}\\
p^-({\bf x})&=&\int_{\partial\mathbb{D}_F} \bar F^{+*}({\bf x},{\bf x}_F)p^+({\bf x}_F){\rm d}^2{\bf x}_F 
+\int_{\partial\mathbb{D}_F} F^-({\bf x},{\bf x}_F)p^-({\bf x}_F){\rm d}^2{\bf x}_F,\label{eq13bbdeco}
\end{eqnarray}
for $x_3\ge x_{3,F}$. These expressions are exact and hold for dissipative media. Making similar substitutions as in section \ref{sec2.4} we obtain
\begin{eqnarray}
2G^{+,f}({\bf x},{\bf x}_S)&=& \int_{\partial\mathbb{D}_F} F^+({\bf x},{\bf x}_F)R({\bf x}_F,{\bf x}_S){\rm d}^2{\bf x}_F 
+\bar F^{-*}({\bf x},{\bf x}_S),\label{eq13Gdeco}\\
2G^{-,f}({\bf x},{\bf x}_S)&=& \int_{\partial\mathbb{D}_F} F^-({\bf x},{\bf x}_F)R({\bf x}_F,{\bf x}_S){\rm d}^2{\bf x}_F 
+\bar F^{+*}({\bf x},{\bf x}_S),\label{eq13Gbbdeco}
\end{eqnarray}
for $x_3\ge x_{3,F}$. Here $G^{\pm,f}({\bf x},{\bf x}_S)$ stands for the downgoing ($+$) and upgoing ($-$) part of the Green's function $G^{p,f}({\bf x},{\bf x}_S)$.

When the medium is lossless and when evanescent waves are neglected at $\partial\mathbb{D}_F$ and at depth level $x_3$ inside the medium, 
the bars on $F^+$ and $F^-$ in representations (\ref{eq13Gdeco}) and (\ref{eq13Gbbdeco}) can be omitted.
Using the Marchenko method, these decomposed focusing functions can then be retrieved from the reflection response $R({\bf x}_F,{\bf x}_S)$ and 
a smooth background model \citep{Wapenaar2013PRL, Slob2014GEO}. 
Once the focusing functions are found, they can be used to retrieve the decomposed Green's functions 
$G^{+,f}({\bf x},{\bf x}_S)$ and $G^{-,f}({\bf x},{\bf x}_S)$ 
(from equations (\ref{eq13Gdeco}) and (\ref{eq13Gbbdeco})) 
and all components of the transfer matrix ${{{\mbox{\boldmath ${\cal T}$}}}}({\bf x},{\bf x}_F)$ (from equation (\ref{eq58vv})).

\section{Conclusions}\label{sec6}

We have derived relations between acoustic propagator matrices, transfer matrices and Marchenko focusing functions.
In the appendices we generalize the expressions for other wave phenomena.
All relations hold for a heterogeneous dissipative medium below a homogeneous upper half-space and account for propagating and evanescent waves.
Only for the transfer matrix beyond the acoustic approach (Appendix \ref{AppB}) we assume that there are no lateral variations at the depth level 
inside the medium where decomposition takes place. 

The derived relations provide insight in the connections between the propagator matrices, transfer matrices and Marchenko focusing functions and may lead to new modelling algorithms for these quantities.
Moreover, several of the derived relations may be useful to develop improved Marchenko-type wave field retrieval and imaging schemes
for different wave phenomena, possibly accounting for evanescent waves inside the medium. 

\section*{Acknowledgements}
We thank Ivan Vasconcelos and an anonymous reviewer for their useful comments, which helped us to improve this paper.
The research of KW has received funding from the European Union's Horizon 2020 research and innovation programme: European Research Council (grant agreement 742703).

\section*{Data Availability}

No data have been used for this study.

\bibliographystyle{gji}


\newpage
\appendix

\section{Unified propagator matrix and focusing functions}\label{AppA}

In this appendix, we extend the theory of section \ref{sec2} to unified wave fields.

\subsection{Unified matrix-vector wave equation}\label{sec4.1}
 
Consider matrix-vector wave equation (\ref{eq2.0}). We partition the $N\times 1$ wave field vector  ${\bf q}$, 
the $N\times 1$ source vector ${\bf d}$ and
the $N\times N$ operator matrix $\mbox{\boldmath ${\cal A}$}$ as follows
\begin{eqnarray}
{\bf q}=\begin{pmatrix}{\bf q}_1\\ {\bf q}_2\end{pmatrix},
\quad{\bf d}=\begin{pmatrix}{\bf d}_1\\ {\bf d}_2\end{pmatrix},
\quad \mbox{\boldmath ${\cal A}$}=\begin{pmatrix}\mbox{\boldmath ${\cal A}$}_{11}&\mbox{\boldmath ${\cal A}$}_{12}\\
\mbox{\boldmath ${\cal A}$}_{21}&\mbox{\boldmath ${\cal A}$}_{22}\end{pmatrix}.
\label{eq7em}
\end{eqnarray}
This includes the acoustic situation (for $N=2$) discussed in section \ref{sec2}.
The vectors and operator matrix for other wave phenomena can be found in various references
\citep{Woodhouse74GJR, Ursin83GEO, Stralen97PHD,  Gelinsky97GEO, Haartsen97JGR, White2006SIAM, Loseth2007GJI}.
A comprehensive overview is given by \citet{Wapenaar2019GJI} for acoustic waves ($N=2$), 
quantum mechanical waves obeying the Schr\"odinger equation ($N=2$), electromagnetic waves ($N=4$),
elastodynamic waves ($N=6$), poroelastodynamic waves ($N=8$), piezoelectric waves ($N=10$) and seismoelectric waves ($N=12$).
For all these wave phenomena, the operator matrix $\mbox{\boldmath ${\cal A}$}$ obeys the following symmetry properties
\begin{eqnarray}
\mbox{\boldmath ${\cal A}$}^t&=&-{\bf N}\mbox{\boldmath ${\cal A}$}{\bf N}^{-1},\label{eqsym1f}\\
\mbox{\boldmath ${\cal A}$}^\dagger&=&-{\bf K}{\,\,\,\bar{\mbox{\!\!\!\boldmath ${\cal A}$}}}{\bf K}^{-1},\label{eqsym2f}\\
\mbox{\boldmath ${\cal A}$}^*&=&{\bf J}{\,\,\,\bar{\mbox{\!\!\!\boldmath ${\cal A}$}}}{\bf J}^{-1},\label{eqsym3f}
\end{eqnarray}
with
\begin{eqnarray}
{\bf N}=\begin{pmatrix} {\bf O} & {\bf I} \\ -{\bf I} & {\bf O}\end{pmatrix},\,
{\bf K}=\begin{pmatrix} {\bf O} & {\bf I} \\ {\bf I} & {\bf O}\end{pmatrix},\,
{\bf J}=\begin{pmatrix} {\bf I}& {\bf O}\\ {\bf O} & -{\bf I}\end{pmatrix},\label{eq23a}
\end{eqnarray}
where ${\bf O}$ and ${\bf I}$ are zero and identity matrices of appropriate size.
Superscript $t$ denotes transposition of the matrix and the operators contained in it, with $\partial_\alpha^t=-\partial_\alpha$. 
Superscript $\dagger$ denotes transposition and complex conjugation. 
The bar on an operator denotes that it is defined in the adjoint medium.
For further details we refer to the aforementioned references.

We show that, with some modifications, equations (\ref{eq7em}) -- (\ref{eqsym3f}) also hold for the matrix-vector form of the Dirac equation for the $4\times 1$ spinor $\mbox{\boldmath $\psi$}$.
The Dirac equation is given by \citep{Book67Sakurai}
\begin{eqnarray}
\mbox{\boldmath $\gamma$}_\mu\partial_\mu\mbox{\boldmath $\psi$} + \frac{mc}{\hbar}\mbox{\boldmath $\psi$}={\bf 0}\label{eqD1}
\end{eqnarray}
(summation over $\mu$ from $1$ to $4$), with the Dirac spinor partitioned as
\begin{eqnarray}
\mbox{\boldmath $\psi$}&=&\begin{pmatrix}\mbox{\boldmath $\psi$}_1\\\mbox{\boldmath $\psi$}_2\end{pmatrix},
\end{eqnarray}
and with
\begin{eqnarray}
\mbox{\boldmath $\gamma$}_k&=&\begin{pmatrix}{\bf O}&-i\mbox{\boldmath $\sigma$}_k\\i\mbox{\boldmath $\sigma$}_k&{\bf O}\end{pmatrix}\quad(k=1,2,3),\label{eqD2}\\
\mbox{\boldmath $\gamma$}_4&=&\begin{pmatrix}{\bf I}&{\bf O}\\{\bf O}&-{\bf I}\end{pmatrix},\label{eqD3}\\
\partial_4&=&\frac{1}{ic}\partial_t+ \frac{eV}{\hbar c},\quad \hbar=\frac{h}{2\pi},
\end{eqnarray}
with 
$V({\bf x})$ the space-dependent potential, 
$h$ Planck's constant, $c$ the speed of light, $e$ the electron charge and $\mbox{\boldmath $\sigma$}_1$, $\mbox{\boldmath $\sigma$}_2$ and $\mbox{\boldmath $\sigma$}_3$ the Pauli matrices, defined as
\begin{eqnarray}
\mbox{\boldmath $\sigma$}_1=\begin{pmatrix} 0 & 1 \\ 1 & 0\end{pmatrix},\quad
\mbox{\boldmath $\sigma$}_2=\begin{pmatrix} 0 & -i \\ i & 0\end{pmatrix},\quad
\mbox{\boldmath $\sigma$}_3=\begin{pmatrix} 1 & 0 \\ 0 & -1\end{pmatrix}.\label{eqPauli}
\end{eqnarray}
Assuming a time-dependence $\exp(-iEt/\hbar)$, we replace $\partial_4$ by $-(E- eV)/\hbar c$.
Equation (\ref{eqD1}) can be rewritten as follows
\begin{eqnarray}
&&\begin{pmatrix}0&-i\mbox{\boldmath $\sigma$}_k\\i\mbox{\boldmath $\sigma$}_k&0\end{pmatrix}\partial_k\begin{pmatrix}\mbox{\boldmath $\psi$}_1\\\mbox{\boldmath $\psi$}_2\end{pmatrix}
-\begin{pmatrix}{\bf I}&{\bf O}\\{\bf O}&-{\bf I}\end{pmatrix}\Bigl(\frac{E-eV}{\hbar c}\Bigl)\begin{pmatrix}\mbox{\boldmath $\psi$}_1\\\mbox{\boldmath $\psi$}_2\end{pmatrix}
+\frac{mc}{\hbar}\begin{pmatrix}\mbox{\boldmath $\psi$}_1\\\mbox{\boldmath $\psi$}_2\end{pmatrix}={\bf 0},
\end{eqnarray}
or
\begin{eqnarray}
&&i\mbox{\boldmath $\sigma$}_k\partial_k\mbox{\boldmath $\psi$}_2+\Bigl(\frac{E-eV-mc^2}{\hbar c}\Bigr)\mbox{\boldmath $\psi$}_1={\bf 0},\\
&&i\mbox{\boldmath $\sigma$}_k\partial_k\mbox{\boldmath $\psi$}_1+\Bigl(\frac{E-eV+mc^2}{\hbar c}\Bigr)\mbox{\boldmath $\psi$}_2={\bf 0}.
\end{eqnarray}
Pre-multiplying all terms by $-i\mbox{\boldmath $\sigma$}_3$ and  bringing the $\partial_3$-operators to the left-hand side, 
using $\mbox{\boldmath $\sigma$}_3\mbox{\boldmath $\sigma$}_1=i\mbox{\boldmath $\sigma$}_2$,
$\mbox{\boldmath $\sigma$}_3\mbox{\boldmath $\sigma$}_2=-i\mbox{\boldmath $\sigma$}_1$ and $\mbox{\boldmath $\sigma$}_3\mbox{\boldmath $\sigma$}_3={\bf I}$,
yields a set of equations which can be recast in the form of equation (\ref{eq2.0}), with $4\times 1$ vectors ${\bf q}$ and ${\bf d}$ and 
 $4\times 4$ operator matrix $\mbox{\boldmath ${\cal A}$}$ partitioned as in equation (\ref{eq7em}), with
\begin{eqnarray}
{\bf q}_1&=&\mbox{\boldmath $\psi$}_1,\quad {\bf q}_2=\mbox{\boldmath $\psi$}_2, \quad {\bf d}_1={\bf d}_2={\bf 0},\\
\mbox{\boldmath ${\cal A}$}_{11}&=&\mbox{\boldmath ${\cal A}$}_{22}=i(\mbox{\boldmath $\sigma$}_1\partial_2-\mbox{\boldmath $\sigma$}_2\partial_1),\label{eq8em}\\
\mbox{\boldmath ${\cal A}$}_{12}&=&i\mbox{\boldmath $\sigma$}_3\Bigl(\frac{E-eV+mc^2}{\hbar c}\Bigr),\label{eq9em}\\
\mbox{\boldmath ${\cal A}$}_{21}&=&i\mbox{\boldmath $\sigma$}_3\Bigl(\frac{E-eV-mc^2}{\hbar c}\Bigr).\label{eq10em}
\end{eqnarray}
With these definitions of the operator submatrices, matrix $\mbox{\boldmath ${\cal A}$}$ obeys symmetry relations (\ref{eqsym1f}) -- (\ref{eqsym3f}), with ${\bf N}$, ${\bf K}$ and ${\bf J}$ defined as
\begin{eqnarray}
{\bf N}=\begin{pmatrix} {\bf O} & i\mbox{\boldmath $\sigma$}_1 \\ i\mbox{\boldmath $\sigma$}_1 & {\bf O}\end{pmatrix},\,
{\bf K}=\begin{pmatrix} {\bf O} & \mbox{\boldmath $\sigma$}_3 \\ \mbox{\boldmath $\sigma$}_3 & {\bf O}\end{pmatrix},\,
{\bf J}=\begin{pmatrix} \mbox{\boldmath $\sigma$}_2& {\bf O}\\ {\bf O} & \mbox{\boldmath $\sigma$}_2\end{pmatrix}.\label{eq23b}
\end{eqnarray}
Although there are no direct applications for geophysics, the Schr\"odinger  and Dirac equations are included in all derivations below, since this comes almost for free. 
When we speak of the `medium', for the Schr\"odinger and Dirac equations it should be understood as the `potential'.

\subsection{Unified propagator matrix}\label{sec4.2}

We define the unified $N\times N$ propagator matrix ${\bf W}({\bf x},{\bf x}_F)$ as the solution of wave equation (\ref{eq2.1}), 
with boundary condition (\ref{eq9998d}) and with the operator matrix $\mbox{\boldmath ${\cal A}$}$ being the unified operator matrix discussed in Appendix \ref{sec4.1}.
Using equation (\ref{eq1330}), the unified wave field vector ${\bf q}({\bf x})$ can be propagated from $x_{3,F}$  to any depth level $x_3$, assuming there are no sources between these depth levels.
We partition ${\bf W}({\bf x},{\bf x}_F)$ as
\begin{eqnarray}\label{eq424b}
{\bf W}({\bf x},{\bf x}_F)= \begin{pmatrix}{\bf W}_{11}      & {\bf W}_{12} \\
                {\bf W}_{21}& {\bf W}_{22}   \end{pmatrix}({\bf x},{\bf x}_F), 
\end{eqnarray}
where ${\bf W}_{11}$, ${\bf W}_{12}$, ${\bf W}_{21}$ and ${\bf W}_{22}$ are $\frac{N}{2}\times\frac{N}{2}$ submatrices of ${\bf W}$.
For each of these submatrices, the second subscript refers to the wave field component (${\bf q}_1$ or ${\bf q}_2$) it acts on at ${\bf x}_F$, whereas the first subscript refers to the wave field component it contributes to at ${\bf x}$.
${\bf W}({\bf x},{\bf x}_F)$ obeys the recursive relation (\ref{eq1330k}), and from equation (\ref{eq1330kinv}) it follows that ${\bf W}({\bf x}_F,{\bf x})$ is the inverse of ${\bf W}({\bf x},{\bf x}_F)$.

\subsection{Relation between unified propagator matrix and Marchenko focusing functions}\label{sec4.3}

We assume again that the medium at and above $\partial\mathbb{D}_F$ is homogeneous and may be dissipative.
The medium below $\partial\mathbb{D}_F$ may be inhomogeneous and dissipative, and it is source-free.
In preparation for defining the focusing functions, in the upper half-space (i.e., at and above $\partial\mathbb{D}_F$) we apply eigenvalue decomposition to matrix  $\tilde{\bf A}$ 
(the spatial  Fourier transform of operator matrix $\mbox{\boldmath ${\cal A}$}$), as follows
\begin{eqnarray}
\tilde{\bf A}=\tilde{\bf L}\tilde{\bf \Lambda}\tilde{\bf L}^{-1},\label{eqA21}
\end{eqnarray}
with
\begin{eqnarray}\label{Aeq7mvbbprffBBrev}
\tilde{{{\mbox{\boldmath $\Lambda$}}}}= \begin{pmatrix} i\omega{\bf S}_3^+    &{\bf O}\\
              {\bf O} & -i\omega{\bf S}_3^-   \end{pmatrix},\quad
\tilde{\bf L}= \begin{pmatrix}\tilde{\bf L}_1^+      & \tilde{\bf L}_1^- \\
              \tilde{\bf L}_2^+ & \tilde{\bf L}_2^-    \end{pmatrix},
\end{eqnarray}
with ${\bf S}_3^+$ and ${\bf S}_3^-$ being diagonal matrices containing vertical slownesses  for downgoing and upgoing waves, respectively. 
We express the Fourier transformed wave field vector $\tilde {\bf q}({\bf s},x_3)$
in terms of downgoing and upgoing waves $\tilde{\bf p}^+({\bf s},x_3)$ and $\tilde{\bf p}^-({\bf s},x_3)$ via
\begin{eqnarray}\label{eq2.10rev}
\tilde {\bf q}({\bf s},x_3)=\tilde{\bf L}({\bf s},x_3) \tilde {\bf p}({\bf s},x_3),
\end{eqnarray}
with
\begin{eqnarray}
 \tilde {\bf p}({\bf s},x_3)&=&\begin{pmatrix}\tilde{\bf p}^+\\ \tilde{\bf p}^-\end{pmatrix}({\bf s},x_3).\label{eq2.11rev}
\end{eqnarray}
Note that these equations imply $\tilde {\bf q}_1= \tilde {\bf L}_1^+\tilde {\bf p}^+ +  \tilde {\bf L}_1^-\tilde {\bf p}^-$.
Similar as in section \ref{sec2.3}, equation (\ref{eqdecomfield}), we continue with downgoing and upgoing waves $\tilde {\bf q}_1^+$ and $\tilde {\bf q}_1^-$ which are field-normalised such that
$\tilde{\bf q}_1=\tilde{\bf q}_1^++\tilde{\bf q}_1^-$. To this end, we define $\tilde {\bf q}_1^\pm=\tilde{\bf L}_1^\pm \tilde {\bf p}^\pm$
and we replace equation (\ref{eq2.10rev}) by
\begin{eqnarray}\label{eq2.7rev}
\tilde {\bf q}({\bf s},x_3)=\tilde{\bf D}({\bf s},x_3) \tilde {\bf b}({\bf s},x_3),
\end{eqnarray}
where
\begin{eqnarray}
\tilde{\bf D}({\bf s},x_3)&=&\begin{pmatrix}{\bf I} &{\bf I}\\ \tilde {\bf D}_1^+& \tilde {\bf D}_1^-\end{pmatrix}({\bf s},x_3),\label{eq2.9rev}\\
\tilde {\bf b}({\bf s},x_3)&=&\begin{pmatrix}\tilde{\bf q}_1^+\\ \tilde{\bf q}_1^-\end{pmatrix}({\bf s},x_3),\label{eq2.8rev}
\end{eqnarray}
with
\begin{eqnarray}\label{eq2.14rev}
\tilde{\bf D}_1^\pm=\tilde{\bf L}_2^\pm (\tilde{\bf L}_1^\pm)^{-1}.
\end{eqnarray}
Whereas there is ambiguity in the normalization of the matrices $\tilde{\bf L}_1^\pm$ and $\tilde{\bf L}_2^\pm$, the matrix $\tilde {\bf D}_1^\pm$ is 
uniquely defined (for each wave phenomenon). 
Note that for the acoustic situation we have $\tilde{\bf L}_1^\pm=1$, hence $\tilde{\bf D}_1^\pm=\tilde{\bf L}_2^\pm$, $\tilde{\bf D}=\tilde{\bf L}$ and $\tilde{\bf b}=\tilde{\bf p}$.
Some other examples of matrix $\tilde {\bf D}_1^\pm$ (for electromagnetic and elastodynamic waves) are given by \cite{Wapenaar2022JASA}.
In Appendix \ref{AppC} we derive for any wave phenomenon 
\begin{eqnarray}\label{eq34rev}
\tilde{\bar{{\bf D}}}_1^\pm({\bf s},x_3)={\bf J}_{22}\{{\tilde{\bf D}}_1^\mp(-{\bf s},x_3)\}^*{\bf J}_{11}^{-1},
\end{eqnarray}
with ${\bf J}_{11}$ and ${\bf J}_{22}$ being the $\frac{N}{2}\times\frac{N}{2}$ submatrices of $N\times N$ matrix ${\bf J}$. From equation (\ref{eq23a}) 
we have for all wave phenomena except for the Dirac equation ${\bf J}_{11}=-{\bf J}_{22}={\bf I}$, and
from equation (\ref{eq23b}) we have for the Dirac equation ${\bf J}_{11}={\bf J}_{22}=\mbox{\boldmath $\sigma$}_2$.

We use equation (\ref{eq2.7rev}) at $\partial\mathbb{D}_F$ and the properties of matrix $\tilde{\bar{{\bf D}}}_1^\pm({\bf s},x_3)$  
to derive unified focusing functions and express them in the components of the unified propagator matrix and vice-versa. 
First we aim to substitute equation (\ref{eq2.7rev}) for $x_3=x_{3,F}$ into a transformed version of equation (\ref{eq1330}). 
This equation contains the propagator matrix ${\bf W}({\bf x},{\bf x}_F)$.
For a function  of two space variables, $u({{\bf x}},{\bf x}_F)$ (with ${\bf x}_F$ at $\partial\mathbb{D}_F$), 
we define the spatial Fourier transformation along the horizontal components of the second space variable as
\begin{eqnarray}
&&\tilde u({{\bf x}},{{\bf s}},x_{3,F})=
\int_{{\mathbb{R}}^2}u({{\bf x}},{\bf x}_{{\rm H},F},x_{3,F})\exp\{i\omega{{\bf s}}\cdot{\bf x}_{{\rm H},F}\}{\rm d}^2{\bf x}_{{\rm H},F}\label{eq999329}
\end{eqnarray}
and its inverse as
\begin{eqnarray}
&&u({{\bf x}},{\bf x}_{{\rm H},F},x_{3,F})=
\frac{\omega^2}{4\pi^2}\int_{{\mathbb{R}}^2}\tilde u({{\bf x}},{{\bf s}},x_{3,F})\exp\{-i\omega{{\bf s}}\cdot{\bf x}_{{\rm H},F}\}{\rm d}^2{\bf s}.\label{eq999329inv}
\end{eqnarray}
Note that the sign in the exponential of equation (\ref{eq999329})  is opposite to that in equation (\ref{eq99950b}).
Using these definitions and Parseval's theorem, we rewrite equation (\ref{eq1330}) as
\begin{eqnarray}
{\bf q}({{\bf x}})=\frac{\omega^2}{4\pi^2}\int_{{\mathbb{R}}^2} \tilde{\bf W}({{\bf x}},{{\bf s}},{x_{3,F}})\tilde{\bf q}({{\bf s}},{x_{3,F}}){\rm d}^2{{\bf s}},
 \label{eq99910}
\end{eqnarray}
with $\tilde{\bf W}({{\bf x}},{{\bf s}},{x_{3,F}})$ obeying the boundary condition 
\begin{eqnarray}
 \tilde{\bf W}({{\bf x}},{{\bf s}},{x_{3,F}})|_{x_3={x_{3,F}}}={\bf I}\exp\{i\omega{{\bf s}}\cdot{{\bf x}_{{\rm H}}}\}.\label{eq99930}
\end{eqnarray}
Substitution of equation (\ref{eq2.7rev}) into equation (\ref{eq99910}) gives
\begin{eqnarray}
{\bf q}({{\bf x}})=\frac{\omega^2}{4\pi^2}\int_{{\mathbb{R}}^2} \tilde{\bf Y}({{\bf x}},{{\bf s}},{x_{3,F}})\tilde{\bf b}({{\bf s}},{x_{3,F}}){\rm d}^2{{\bf s}}
 \label{eq99910y}
\end{eqnarray}
for $x_3\ge x_{3,F}$, with
\begin{eqnarray}
 \tilde{\bf Y}({{\bf x}},{{\bf s}},{x_{3,F}})=\tilde{\bf W}({{\bf x}},{{\bf s}},{x_{3,F}})\tilde{\bf D}({{\bf s}},x_{3,F}).\label{eq99943y}
 \end{eqnarray}
We partition matrix $\tilde{\bf Y}({{\bf x}},{{\bf s}},{x_{3,F}})$ as follows
\begin{eqnarray}\label{eq2.20}
\tilde{\bf Y}({{\bf x}},{{\bf s}},{x_{3,F}})=\begin{pmatrix}\tilde {\bf Y}_1^+ & \tilde {\bf Y}_1^-\\ \tilde {\bf Y}_2^+ & \tilde {\bf Y}_2^-\end{pmatrix}({{\bf x}},{{\bf s}},{x_{3,F}}).
\end{eqnarray}
Using equation  (\ref{eq2.9rev}) and the spatial Fourier transformation of equation (\ref{eq424b}), we obtain
\begin{eqnarray}
&&\tilde{\bf Y}_1^\pm({{\bf x}},{{\bf s}},{x_{3,F}})=
\tilde{\bf W}_{11}({{\bf x}},{{\bf s}},{x_{3,F}})+\tilde{\bf W}_{12}({{\bf x}},{{\bf s}},{x_{3,F}})\tilde{\bf D}_1^\pm({\bf s},x_{3,F}),\label{eq43}\\
&&\tilde{\bf Y}_2^\pm({{\bf x}},{{\bf s}},{x_{3,F}})=
\tilde{\bf W}_{21}({{\bf x}},{{\bf s}},{x_{3,F}})+\tilde{\bf W}_{22}({{\bf x}},{{\bf s}},{x_{3,F}})\tilde{\bf D}_1^\pm({\bf s},x_{3,F}).\label{eq44}
\end{eqnarray}
We analyse these expressions one by one. First consider $\tilde{\bf Y}_1^-({{\bf x}},{{\bf s}},{x_{3,F}})$. Via equation (\ref{eq99910y}) 
it can be seen that subscript $1$ refers to wavefield component ${\bf q}_1$ at ${\bf x}$ and superscript
$-$ refers to the upgoing wavefield component  $\tilde {\bf q}_1^-$ at $x_{3,F}$. Moreover, for $x_3=x_{3,F}$ we obtain, using equation
(\ref{eq99930}), $\tilde{\bf Y}_1^-({{\bf x}},{{\bf s}},{x_{3,F}})|_{x_3={x_{3,F}}}={\bf I}\exp\{i\omega{{\bf s}}\cdot{{\bf x}_{{\rm H}}}\}$, 
or, applying the inverse spatial Fourier transformation defined in equation (\ref{eq999329inv}),
${\bf Y}_1^-({\bf x},{\bf x}_F)|_{x_3={x_{3,F}}} = {\bf I}\delta({{\bf x}_{\rm H}}-{{\bf x}_{{\rm H},F}})$,
which is a focusing condition. Hence, we define
\begin{eqnarray}
\tilde{\bf Y}_1^-({{\bf x}},{{\bf s}},{x_{3,F}})=\tilde{\bf F}_1({{\bf x}},{{\bf s}},{x_{3,F}})
=\tilde{\bf W}_{11}({{\bf x}},{{\bf s}},{x_{3,F}})+\tilde{\bf W}_{12}({{\bf x}},{{\bf s}},{x_{3,F}})\tilde{\bf D}_1^-({\bf s},x_{3,F}),\label{eqy1min}
\end{eqnarray}
with $\tilde{\bf F}_1({{\bf x}},{{\bf s}},{x_{3,F}})$ denoting the spatial Fourier transform of the focusing function ${\bf F}_1({\bf x},{\bf x}_F)$ 
for wavefield component ${\bf q}_1$, which focuses as an upgoing field 
 at ${\bf x}={\bf x}_F$ and continues as an upgoing field in the homogeneous upper half-space. 
Note that the focusing function is a $\frac{N}{2}\times\frac{N}{2}$ matrix.
Next, we consider $\tilde{\bf Y}_2^-({{\bf x}},{{\bf s}},{x_{3,F}})$. Subscript $2$ refers to wavefield component ${\bf q}_2$ at ${\bf x}$ and superscript
$-$ refers again to the upgoing wavefield component  $\tilde {\bf q}_1^-$ at $x_{3,F}$. For $x_3=x_{3,F}$ we obtain, using equation (\ref{eq99930}),
$\tilde{\bf Y}_2^-({{\bf x}},{{\bf s}},{x_{3,F}})|_{x_3={x_{3,F}}}=\tilde{\bf D}_1^-({\bf s},x_{3,F})\exp\{i\omega{{\bf s}}\cdot{{\bf x}_{{\rm H}}}\}$, which is a focusing condition,
but somewhat more complicated than for $\tilde{\bf Y}_1^-({{\bf x}},{{\bf s}},{x_{3,F}})$ because of the mix of wavefield components ${\bf q}_2$ and $\tilde {\bf q}_1^-$. 
Hence, we define
\begin{eqnarray}
\tilde{\bf Y}_2^-({{\bf x}},{{\bf s}},{x_{3,F}})=\tilde{\bf F}_2({{\bf x}},{{\bf s}},{x_{3,F}})=
\tilde{\bf W}_{21}({{\bf x}},{{\bf s}},{x_{3,F}})+\tilde{\bf W}_{22}({{\bf x}},{{\bf s}},{x_{3,F}})\tilde{\bf D}_1^-({\bf s},x_{3,F}),\label{eqy2min}
\end{eqnarray}
with $\tilde{\bf F}_2({{\bf x}},{{\bf s}},{x_{3,F}})$ denoting the spatial
Fourier transform of the focusing function ${\bf F}_2({\bf x},{\bf x}_F)$ for wavefield component ${\bf q}_2$, 
which focuses as an upgoing field at ${\bf x}={\bf x}_F$ and continues as an upgoing field in the homogeneous upper half-space
(note that the definition of $\tilde{\bf F}_2$ is different from that in \cite{Wapenaar2022JASA}, to facilitate the derivations below).
The focusing functions $\tilde{\bf F}_1({{\bf x}},{{\bf s}},{x_{3,F}})$ and $\tilde{\bf F}_2({{\bf x}},{{\bf s}},{x_{3,F}})$ together form the right column of matrix $ \tilde{\bf Y}({{\bf x}},{{\bf s}},{x_{3,F}})$.

For the analysis of the submatrices in the left column of $\tilde{\bf Y}({{\bf x}},{{\bf s}},{x_{3,F}})$, we use symmetry relation (\ref{eq34rev}) and we need a similar relation for the 
submatrices of $\tilde{\bf W}({{\bf x}},{{\bf s}},{x_{3,F}})$.
In Appendix \ref{AppC} we derive
\begin{eqnarray}\label{eq65awss}
\bar{\bf W}({\bf x},{\bf x}_F)={\bf J}{\bf W}^*({\bf x},{\bf x}_F){\bf J}^{-1}.
\end{eqnarray}
From the spatial Fourier transform of this equation we obtain for the submatrices of $\tilde{\bf W}({{\bf x}},{{\bf s}},{x_{3,F}})$
\begin{eqnarray}\label{eq25}
\tilde{\bar{{\bf W}}}_{\alpha\beta}({{\bf x}},{{\bf s}},{x_{3,F}})={\bf J}_{\alpha\alpha}{\tilde{\bf W}}_{\alpha\beta}^*({{\bf x}},-{{\bf s}},{x_{3,F}}){\bf J}_{\beta\beta}^{-1}
\end{eqnarray}
(no summation for repeated subscripts). 
Substituting equations (\ref{eq34rev}) and (\ref{eq25}) into equations (\ref{eq43}) and (\ref{eq44}) yields
\begin{eqnarray}
\tilde{\bar{{\bf Y}}}_1^+({{\bf x}},{{\bf s}},{x_{3,F}})&=&{\bf J}_{11}\tilde{\bf Y}_1^{-*}({{\bf x}},-{{\bf s}},{x_{3,F}}){\bf J}_{11}^{-1},\\
\tilde{\bar{{\bf Y}}}_2^+({{\bf x}},{{\bf s}},{x_{3,F}})&=&{\bf J}_{22}\tilde{\bf Y}_2^{-*}({{\bf x}},-{{\bf s}},{x_{3,F}}){\bf J}_{11}^{-1}.
\end{eqnarray}
Hence, using equations (\ref{eqy1min}) and (\ref{eqy2min}), we find for the submatrices in the left column of $\tilde{\bf Y}({{\bf x}},{{\bf s}},{x_{3,F}})$
\begin{eqnarray}
\tilde{{\bf Y}}_1^+({{\bf x}},{{\bf s}},{x_{3,F}})&=&{\bf J}_{11}{\tilde{\bar{\bf F}}}_1^*({{\bf x}},-{{\bf s}},{x_{3,F}}){\bf J}_{11}^{-1},\label{eq96k}\\
\tilde{{\bf Y}}_2^+({{\bf x}},{{\bf s}},{x_{3,F}})&=&{\bf J}_{22}{\tilde{\bar{\bf F}}}_2^*({{\bf x}},-{{\bf s}},{x_{3,F}}){\bf J}_{11}^{-1}.\label{eq97k}
\end{eqnarray}
Hence, matrix $\tilde{\bf Y}({{\bf x}},{{\bf s}},{x_{3,F}})$ becomes
\begin{eqnarray}\label{eq67}
&&\tilde{\bf Y}({{\bf x}},{{\bf s}},{x_{3,F}})=\begin{pmatrix} {\bf J}_{11}{\tilde{\bar{\bf F}}}_1^*({{\bf x}},-{{\bf s}},{x_{3,F}}){\bf J}_{11}^{-1} &\tilde{\bf F}_1({{\bf x}},{{\bf s}},{x_{3,F}})\\ 
{\bf J}_{22}{\tilde{\bar{\bf F}}}_2^*({{\bf x}},-{{\bf s}},{x_{3,F}}){\bf J}_{11}^{-1} &\tilde{\bf F}_2({{\bf x}},{{\bf s}},{x_{3,F}})\end{pmatrix},
\end{eqnarray}
or, using the inverse Fourier transformation defined in equation (\ref{eq999329inv}),
\begin{eqnarray}\label{eq66}
{\bf Y}({\bf x},{\bf x}_F)=\begin{pmatrix} {\bf J}_{11}{{\bar{\bf F}}}_1^*({\bf x},{\bf x}_F){\bf J}_{11}^{-1} &{\bf F}_1({\bf x},{\bf x}_F)\\ 
{\bf J}_{22}{{\bar{\bf F}}}_2^*({\bf x},{\bf x}_F){\bf J}_{11}^{-1} &{\bf F}_2({\bf x},{\bf x}_F)\end{pmatrix}.
\end{eqnarray}
This is a generalisation of equation (\ref{eq21}).
Note that $\tilde{\bf F}_1$, $\tilde{\bf F}_2$, ${\tilde{\bar{\bf F}}}_1^*$ and ${\tilde{\bar{\bf F}}}_2^*$ are expressed in terms of the 
submatrices of the propagator matrix $\tilde{\bf W}({{\bf x}},{{\bf s}},{x_{3,F}})$ via equations (\ref{eqy1min}), (\ref{eqy2min}), (\ref{eq96k}) and (\ref{eq97k}). Conversely,
we can express the submatrices of the propagator matrix $\tilde{\bf W}({{\bf x}},{{\bf s}},{x_{3,F}})$ 
in terms of the focusing functions $\tilde{\bf F}_1$, $\tilde{\bf F}_2$, ${\tilde{\bar{\bf F}}}_1^*$ and ${\tilde{\bar{\bf F}}}_2^*$. To this end, we start with inverting equation (\ref{eq99943y}), according to
\begin{eqnarray}\label{eq62}
\tilde{\bf W}({{\bf x}},{{\bf s}},{x_{3,F}}) =\tilde{\bf Y}({{\bf x}},{{\bf s}},{x_{3,F}}) \{\tilde{\bf D}({\bf s},x_{3,F})\}^{-1},
\end{eqnarray}
with
\begin{eqnarray}\label{eq39}
\{\tilde{\bf D}({\bf s},x_3)\}^{-1}=\begin{pmatrix}-(\tilde{\bf \Delta}_1)^{-1}\tilde{\bf D}_1^-& (\tilde{\bf \Delta}_1)^{-1}\\
(\tilde{\bf \Delta}_1)^{-1}\tilde{\bf D}_1^+ & -(\tilde{\bf \Delta}_1)^{-1} \end{pmatrix}({\bf s},x_3),
\end{eqnarray}
\begin{eqnarray}
\tilde{\bf \Delta}_1=\tilde{\bf D}_1^+- \tilde{\bf D}_1^-.\label{eq2107}
\end{eqnarray}
Using equations (\ref{eq67}) and (\ref{eq39}), we obtain
\begin{eqnarray}
\tilde{\bf W}_{\alpha 1}({{\bf x}},{{\bf s}},{x_{3,F}})&=&
-{\bf J}_{\alpha\alpha}{\tilde{\bar{\bf F}}}_\alpha^*({{\bf x}},-{{\bf s}},{x_{3,F}}){\bf J}_{11}^{-1}\{\tilde{\bf \Delta}_1({\bf s},x_{3,F})\}^{-1}\tilde{\bf D}_1^-({\bf s},x_{3,F})
\nonumber\\
&+&\tilde{\bf F}_\alpha({{\bf x}},{{\bf s}},{x_{3,F}})\{\tilde{\bf \Delta}_1({\bf s},x_{3,F})\}^{-1}\tilde{\bf D}_1^+({\bf s},x_{3,F}),\label{eqWa1}\\
\tilde{\bf W}_{\alpha 2}({{\bf x}},{{\bf s}},{x_{3,F}})&=&
{\bf J}_{\alpha\alpha}{\tilde{\bar{\bf F}}}_\alpha^*({{\bf x}},-{{\bf s}},{x_{3,F}}){\bf J}_{11}^{-1}\{\tilde{\bf \Delta}_1({\bf s},x_{3,F})\}^{-1}
\nonumber\\
&-&\tilde{\bf F}_\alpha({{\bf x}},{{\bf s}},{x_{3,F}})\{\tilde{\bf \Delta}_1({\bf s},x_{3,F})\}^{-1}\label{eqWa2}
\end{eqnarray}
(no summation for repeated subscripts). These expressions are a generalisation of equations (\ref{eq14}) -- (\ref{eq15v}). Those equations follow as a special case from equations (\ref{eqWa1}) and (\ref{eqWa2})
by substituting ${\bf J}_{11}=-{\bf J}_{22}=1$, $\tilde {\bf D}_1^\pm({{\bf s}},x_{3,F}) = \pm{s_{3,0}}/{\rho_0}$, $\{\tilde{\bf \Delta}_1({{\bf s}},x_{3,F})\}^{-1}={\rho_0}/{2s_{3,0}}$, 
and applying an inverse spatial Fourier transformation, which involves replacing $s_{3,0}$ by operator $\frac{1}{\omega}{\cal H}_1({\bf x}_F)$.

\subsection{Representations with unified Marchenko focusing functions}\label{sec4.4}

Applying Parseval's theorem to equation (\ref{eq99910y}) and substituting the expressions for ${\bf q}({\bf x})$, ${\bf b}({\bf x}_F)$ and ${\bf Y}({\bf x},{\bf x}_F)$ 
(equations (\ref{eq7em}), (\ref{eq2.8rev}) and (\ref{eq66})), gives
the following representation for the quantities ${\bf q}_1({\bf x})$ and ${\bf q}_2({\bf x})$ inside the inhomogeneous medium 
\begin{eqnarray}
{\bf q}_\alpha({\bf x})&=&\int_{\partial\mathbb{D}_F} {\bf J}_{\alpha\alpha}{{\bar{\bf F}}}_\alpha^*({\bf x},{\bf x}_F){\bf J}_{11}^{-1}{\bf q}_1^+({\bf x}_F){\rm d}^2{\bf x}_F 
+\int_{\partial\mathbb{D}_F} {\bf F}_\alpha({\bf x},{\bf x}_F){\bf q}_1^-({\bf x}_F){\rm d}^2{\bf x}_F\label{eq13gen}
\end{eqnarray}
(no summation for repeated subscripts) for $x_3\ge x_{3,F}$. This is a generalisation of equations (\ref{eq13}) and (\ref{eq13bb}).

We use equation (\ref{eq13gen}) to derive representations for Green's functions between the boundary $\partial\mathbb{D}_F$ and any position ${\bf x}$ inside the medium.
We define a unit ${\bf d}_2$-type source (see equation (\ref{eq7em})) at ${\bf x}_S$ just above $\partial\mathbb{D}_F$. The $\frac{N}{2}\times\frac{N}{2}$ Green's matrix
${\bf G}_{12}({\bf x},{\bf x}_S)$ stands for the ${\bf q}_1$-type field observed at ${\bf x}$, in response to this source.
The spatial Fourier transform of the downgoing component at $\partial\mathbb{D}_F$ (i.e., just below the source) is proportional to the upper-right submatrix of the decomposition operator of
equation (\ref{eq39}), according to
\begin{eqnarray}
\tilde{\bf G}_{12}^+({\bf x}_F,{\bf s},x_{3,S})
=\{\tilde{\bf \Delta}_1({{\bf s}},x_{3,F})\}^{-1}\exp\{i\omega{\bf s}\cdot{\bf x}_{{\rm H},F}\}\label{eq330k}
\end{eqnarray}
\citep{Wapenaar2022JASA}. To compensate for the effects of the inverse matrix $\{\tilde{\bf \Delta}_1({{\bf s}},x_{3,F})\}^{-1}$, we define a modified Green's matrix as
\begin{eqnarray}
\tilde{\bf \Gamma}_{12}({\bf x},{\bf s},x_{3,S})=\tilde{\bf G}_{12}({\bf x},{\bf s},x_{3,S})\tilde{\bf \Delta}_1({\bf s}, x_{3,F}),\label{eq331}
\end{eqnarray}
such that its downgoing component at $\partial\mathbb{D}_F$ is given by
\begin{eqnarray}
\tilde{\bf \Gamma}_{12}^+ ({\bf x}_F,{\bf s},x_{3,S})
={\bf I}\exp\{i\omega{\bf s}\cdot{\bf x}_{{\rm H},F}\},\label{eq544}
\end{eqnarray}
or, after applying an inverse spatial Fourier transformation 
\begin{eqnarray}
{\bf \Gamma}_{12}^+({\bf x}_F,{\bf x}_S)&=&{\bf I}\delta({\bf x}_{{\rm H},F}-{\bf x}_{{\rm H},S}).\label{eq4Gag}
\end{eqnarray}
The upgoing response at $\partial\mathbb{D}_F$ to this downgoing source field is by definition the reflection response, hence
\begin{eqnarray}
{\bf \Gamma}_{12}^-({\bf x}_F,{\bf x}_S)&=&{\bf R}({\bf x}_F,{\bf x}_S).\label{eq4Gagd}
\end{eqnarray}
The field at ${\bf x}$ inside the medium consists of ${\bf \Gamma}_{12}({\bf x},{\bf x}_S)$ and ${\bf \Gamma}_{22}({\bf x},{\bf x}_S)$,
where $\tilde{\bf \Gamma}_{22}({\bf x},{\bf s},x_{3,S})=\tilde{\bf G}_{22}({\bf x},{\bf s},x_{3,S})\tilde{\bf \Delta}_1({\bf s}, x_{3,F})$, with $\tilde{\bf G}_{22}({\bf x},{\bf s},x_{3,S})$ being
the Green's function for the ${\bf q}_2$-type field observed at ${\bf x}$.
Substituting ${\bf q}_\alpha({\bf x})={\bf \Gamma}_{\alpha 2}({\bf x},{\bf x}_S)$ and ${\bf q}_1^\pm({\bf x}_F)={\bf \Gamma}_{12}^\pm({\bf x}_F,{\bf x}_S)$ 
into equation (\ref{eq13gen}), using equations (\ref{eq4Gag}) and (\ref{eq4Gagd}), we obtain
\begin{eqnarray}
{\bf \Gamma}_{\alpha 2}({\bf x},{\bf x}_S)&=&
\int_{\partial\mathbb{D}_F} {\bf F}_\alpha({\bf x},{\bf x}_F){\bf R}({\bf x}_F,{\bf x}_S){\rm d}^2{\bf x}_F
+{\bf J}_{\alpha\alpha}{{\bar{\bf F}}}_\alpha^*({\bf x},{\bf x}_S){\bf J}_{11}^{-1}, \label{eq13genG}
\end{eqnarray}
(no summation for repeated subscripts) for $x_3\ge x_{3,F}$. This is a generalisation of equations (\ref{eq13G}) and (\ref{eq13Gbb}) and a starting point for developing a unified Marchenko method
for full wave fields,  accounting for evanescent waves inside the medium. 
Once the focusing functions are found, they can be used to retrieve the Green's matrices
${\bf \Gamma}_{\alpha 2}({\bf x},{\bf x}_S)$ for $\alpha=1,2$
(from equation (\ref{eq13genG})) and all components of the propagator matrix ${\bf W}({\bf x},{\bf x}_F)$ (from equations (\ref{eqWa1}) and (\ref{eqWa2})).

\section{Unified transfer matrix and decomposed focusing functions}\label{AppB}

In this appendix, we extend the theory of section \ref{sec3} to unified wave fields.

\subsection{Unified transfer matrix}

We introduce the unified transfer matrix as follows.
Given the downgoing and upgoing fields ${\bf q}^+({\bf x}_F)$ and ${\bf q}^-({\bf x}_F)$ contained in vector ${\bf b}({\bf x}_F)$ at the boundary $\partial\mathbb{D}_F$, 
we transfer these fields to depth level $x_3$ via 
\begin{eqnarray}
{\bf b}({\bf x})=\int_{\partial\mathbb{D}_F} {{{\mbox{\boldmath ${\cal T}$}}}}({\bf x},{\bf x}_F){\bf b}({\bf x}_F){\rm d}^2{\bf x}_F,\label{eq1330transB}
\end{eqnarray}
for $x_3\ge x_{3,F}$.
The unified transfer matrix ${{{\mbox{\boldmath ${\cal T}$}}}}({\bf x},{\bf x}_F)$ is partitioned as follows
\begin{eqnarray}\label{eq424TU}
{{{\mbox{\boldmath ${\cal T}$}}}}({\bf x},{\bf x}_F)= \begin{pmatrix}{{{\mbox{\boldmath ${\cal T}$}}}}^{+,+}      & {{{\mbox{\boldmath ${\cal T}$}}}}^{+,-} \\
                {{{\mbox{\boldmath ${\cal T}$}}}}^{-,+} & {{{\mbox{\boldmath ${\cal T}$}}}}^{-,-}    \end{pmatrix}({\bf x},{\bf x}_F),
\end{eqnarray}
with ${{{\mbox{\boldmath ${\cal T}$}}}}^{\pm,\pm}$ being $\frac{N}{2}\times\frac{N}{2}$ submatrices. 
Analogous to equation (\ref{eq31}), matrix ${{{\mbox{\boldmath ${\cal T}$}}}}({\bf x},{\bf x}_F)$ is related to the unified propagator matrix ${\bf W}({\bf x},{\bf x}_F)$ of equation (\ref{eq424b}) via 
\begin{eqnarray}
{{{\mbox{\boldmath ${\cal T}$}}}}({\bf x},{\bf x}_F)={{{\mbox{\boldmath ${\cal D}$}}}}^{-1}({\bf x}){\bf W}({\bf x},{\bf x}_F){{{{\mbox{\boldmath ${\cal D}$}}}}}({\bf x}_F),\label{eq31B}
\end{eqnarray}
with ${{{{\mbox{\boldmath ${\cal D}$}}}}}({\bf x}_F)$ and ${{{\mbox{\boldmath ${\cal D}$}}}}^{-1}({\bf x})$ being 
 the inverse spatial Fourier transforms of $\tilde{\bf D}({\bf s},x_{3,F})$ and $\{\tilde{\bf D}({\bf s},x_3)\}^{-1}$, defined in equations (\ref{eq2.9rev}) and (\ref{eq39}), respectively.
Unlike in the acoustic situation, where ${{{\mbox{\boldmath ${\cal L}$}}}}({\bf x}_F)$ and ${{{\mbox{\boldmath ${\cal L}$}}}}^{-1}({\bf x})$ in equation (\ref{eq31}) account 
for lateral variations of the medium parameters, the unified  matrices $\tilde{\bf D}({\bf s},x_{3,F})$ and $\{\tilde{\bf D}({\bf s},x_3)\}^{-1}$
are defined for laterally invariant medium parameters at depths $x_{3,F}$ and $x_3$.
For $\tilde{\bf D}({\bf s},x_{3,F})$ this is not a restriction, since $x_{3,F}$ is the depth of the boundary $\partial\mathbb{D}_F$ 
between the inhomogeneous medium and the homogeneous upper half-space.
However, for $\{\tilde{\bf D}({\bf s},x_3)\}^{-1}$ it implies that this operator can only be applied at depths where no lateral variations occur. 

\subsection{Relation between unified transfer matrix and decomposed Marchenko focusing functions}

Assuming there are no lateral variations at a specific depth level $x_3$, we use the spatial Fourier transformation defined in equation (\ref{eq99950b}) 
along the horizontal components of the first space variable to express the transfer matrix (analogous to equation (\ref{eq31k})) as
\begin{eqnarray}\label{eq62b}
\tilde{{{\mbox{\boldmath ${\cal T}$}}}}({\bf s},x_3,{\bf x}_F) = \{\tilde{\bf D}({\bf s},x_3)\}^{-1}\tilde{\bf Y}({\bf s},x_3,{\bf x}_F),
\end{eqnarray}
with $\{\tilde{\bf D}({\bf s},x_3)\}^{-1}$ defined in equation (\ref{eq39}) and $\tilde{\bf Y}({\bf s},x_3,{\bf x}_F)$ being the Fourier transform of 
${\bf Y}({\bf x},{\bf x}_F)$ defined in equation (\ref{eq66}).
Analogous to equation (\ref{eq55b}), we obtain for the right column of $\tilde{{{\mbox{\boldmath ${\cal T}$}}}}({\bf s},x_3,{\bf x}_F)$
\begin{eqnarray}\label{eq55}
&&\begin{pmatrix}\tilde{\bf F}^+({\bf s},x_3,{\bf x}_F)\\ \tilde{\bf F}^-({\bf s},x_3,{\bf x}_F)\end{pmatrix}=
\begin{pmatrix}-(\tilde{\bf \Delta}_1)^{-1}\tilde{\bf D}_1^-& (\tilde{\bf \Delta}_1)^{-1}\\
(\tilde{\bf \Delta}_1)^{-1}\tilde{\bf D}_1^+ & -(\tilde{\bf \Delta}_1)^{-1} \end{pmatrix}({\bf s},x_3)
\begin{pmatrix}\tilde{\bf F}_1({\bf s},x_3,{\bf x}_F)\\\tilde{\bf F}_2({\bf s},x_3,{\bf x}_F)\end{pmatrix},
\end{eqnarray}
with $\tilde{\bf F}^+({\bf s},x_3,{\bf x}_F)$ and $\tilde{\bf F}^-({\bf s},x_3,{\bf x}_F)$ being the downgoing and upgoing parts at $x_3$ of $\tilde{\bf F}_1({\bf s},x_3,{\bf x}_F)$.
For the left column of $\tilde{{{\mbox{\boldmath ${\cal T}$}}}}({\bf s},x_3,{\bf x}_F)$ we analyse the following expression
\begin{eqnarray}\label{eq56}
&&\begin{pmatrix}-(\tilde{\bf \Delta}_1)^{-1}\tilde{\bf D}_1^-& (\tilde{\bf \Delta}_1)^{-1}\\
(\tilde{\bf \Delta}_1)^{-1}\tilde{\bf D}_1^+ & -(\tilde{\bf \Delta}_1)^{-1} \end{pmatrix}({\bf s},x_3)
\begin{pmatrix}{\bf J}_{11}{\tilde{\bar{\bf F}}}_1^*(-{\bf s},x_3,{\bf x}_F){\bf J}_{11}^{-1}\\{\bf J}_{22}{\tilde{\bar{\bf F}}}_2^*(-{\bf s},x_3,{\bf x}_F){\bf J}_{11}^{-1}\end{pmatrix}.
\end{eqnarray}
Using  equations (\ref{eq34rev}) and (\ref{eqC16}) in equation (\ref{eq56}) gives
\begin{eqnarray}\label{eq58}
&&\begin{pmatrix}{\bf J}_{11}({\tilde{\bar{\bf \Delta}}}_1^*)^{-1}({\tilde{\bar{\bf D}}}_1^+)^*{\bf J}_{11}^{-1}& -{\bf J}_{11}({\tilde{\bar{\bf \Delta}}}_1^*)^{-1}{\bf J}_{22}^{-1}\\
-{\bf J}_{11}({\tilde{\bar{\bf \Delta}}}_1^*)^{-1}({\tilde{\bar{\bf D}}}_1^-)^*{\bf J}_{11}^{-1} & {\bf J}_{11}({\tilde{\bar{\bf \Delta}}}_1^*)^{-1}{\bf J}_{22}^{-1} \end{pmatrix}(-{\bf s},x_3)
\begin{pmatrix}{\bf J}_{11}{\tilde{\bar{\bf F}}}_1^*(-{\bf s},x_3,{\bf x}_F){\bf J}_{11}^{-1}\\{\bf J}_{22}{\tilde{\bar{\bf F}}}_2^*(-{\bf s},x_3,{\bf x}_F){\bf J}_{11}^{-1}\end{pmatrix}.
\end{eqnarray}
By comparing this with equation (\ref{eq55}) we find that the expression in equation (\ref{eq58}) is equal to
\begin{eqnarray}\label{eq59}
&&\begin{pmatrix}{\bf J}_{11}\{{\tilde{\bar{\bf F}}}^-(-{\bf s},x_3,{\bf x}_F)\}^*{\bf J}_{11}^{-1}\\{\bf J}_{11}\{{\tilde{\bar{\bf F}}}^+(-{\bf s},x_3,{\bf x}_F)\}^*{\bf J}_{11}^{-1}\end{pmatrix}.
\end{eqnarray}
Combining the right column (equation (\ref{eq55})) and the left column (equation (\ref{eq59})), we obtain the following expression for the unified transfer matrix
\begin{eqnarray}\label{eq60}
&&\tilde{{{\mbox{\boldmath ${\cal T}$}}}}({\bf s},x_3,{\bf x}_F)=
\begin{pmatrix} {\bf J}_{11}\{{\tilde{\bar{\bf F}}}^-(-{\bf s},x_3,{\bf x}_F)\}^*{\bf J}_{11}^{-1}&\tilde{\bf F}^+({\bf s},x_3,{\bf x}_F)\\ 
{\bf J}_{11}\{{\tilde{\bar{\bf F}}}^+(-{\bf s},x_3,{\bf x}_F)\}^*{\bf J}_{11}^{-1}&\tilde{\bf F}^-({\bf s},x_3,{\bf x}_F)\end{pmatrix},
\end{eqnarray}
or, in the space domain,
\begin{eqnarray}\label{eq60b}
&&{{{\mbox{\boldmath ${\cal T}$}}}}({\bf x},{\bf x}_F)=
\begin{pmatrix} {\bf J}_{11}\{{{\bar{\bf F}}}^-({\bf x},{\bf x}_F)\}^*{\bf J}_{11}^{-1}&{\bf F}^+({\bf x},{\bf x}_F)\\ 
{\bf J}_{11}\{{{\bar{\bf F}}}^+({\bf x},{\bf x}_F)\}^*{\bf J}_{11}^{-1}&{\bf F}^-({\bf x},{\bf x}_F)\end{pmatrix}.
\end{eqnarray}
This is the generalisation of equation (\ref{eq58vv}).

\subsection{Representations with decomposed unified Marchenko focusing functions}

Substituting the expressions for ${\bf b}({\bf x})$ and ${{{\mbox{\boldmath ${\cal T}$}}}}({\bf x},{\bf x}_F)$ 
into equation (\ref{eq1330transB}) gives
the following representations for the downgoing and upgoing fields, ${\bf q}_1^+({\bf x})$ and ${\bf q}_1^-({\bf x})$ respectively, inside the inhomogeneous medium 
\begin{eqnarray}
{\bf q}_1^+({\bf x})&=&\int_{\partial\mathbb{D}_F} {\bf J}_{11}\{{{\bar{\bf F}}}^-({\bf x},{\bf x}_F)\}^*{\bf J}_{11}^{-1}{\bf q}_1^+({\bf x}_F){\rm d}^2{\bf x}_F 
+\int_{\partial\mathbb{D}_F} {\bf F}^+({\bf x},{\bf x}_F){\bf q}_1^-({\bf x}_F){\rm d}^2{\bf x}_F,\label{eq13decoun}\\
{\bf q}_1^-({\bf x})&=&\int_{\partial\mathbb{D}_F} {\bf J}_{11}\{{{\bar{\bf F}}}^+({\bf x},{\bf x}_F)\}^*{\bf J}_{11}^{-1}{\bf q}_1^+({\bf x}_F){\rm d}^2{\bf x}_F 
+\int_{\partial\mathbb{D}_F} {\bf F}^-({\bf x},{\bf x}_F){\bf q}_1^-({\bf x}_F){\rm d}^2{\bf x}_F,\label{eq13bbdecoun}
\end{eqnarray}
for $x_3\ge x_{3,F}$. These expressions are exact and hold for dissipative media. Making similar substitutions as in section \ref{sec4.4} we obtain
\begin{eqnarray}
{\bf \Gamma}_{12}^+({\bf x},{\bf x}_S)&=&
\int_{\partial\mathbb{D}_F} {\bf F}^+({\bf x},{\bf x}_F){\bf R}({\bf x}_F,{\bf x}_S){\rm d}^2{\bf x}_F
+{\bf J}_{11}\{{{\bar{\bf F}}}^-({\bf x},{\bf x}_S)\}^*{\bf J}_{11}^{-1}, \label{eq13genGdeco}\\
{\bf \Gamma}_{12}^-({\bf x},{\bf x}_S)&=&
\int_{\partial\mathbb{D}_F} {\bf F}^-({\bf x},{\bf x}_F){\bf R}({\bf x}_F,{\bf x}_S){\rm d}^2{\bf x}_F
+{\bf J}_{11}\{{{\bar{\bf F}}}^+({\bf x},{\bf x}_S)\}^*{\bf J}_{11}^{-1}, \label{eq13genGdecoup}
\end{eqnarray}
for $x_3\ge x_{3,F}$. Here ${\bf \Gamma}_{12}^+({\bf x},{\bf x}_S)$ and ${\bf \Gamma}_{12}^-({\bf x},{\bf x}_S)$ stand for the downgoing and upgoing part of the
Green's function ${\bf \Gamma}_{12}({\bf x},{\bf x}_S)$. 
These equations are generalisations of equations (\ref{eq13Gdeco}) and (\ref{eq13Gbbdeco}) and
 form a starting point for developing a unified Marchenko method for decomposed wave fields.
Once the focusing functions are found, they can be used to retrieve the decomposed Green's functions 
${\bf \Gamma}_{12}^+({\bf x},{\bf x}_S)$ and ${\bf \Gamma}_{12}^-({\bf x},{\bf x}_S)$ 
(from equations (\ref{eq13genGdeco}) and (\ref{eq13genGdecoup})) 
and all components of the transfer matrix ${{{\mbox{\boldmath ${\cal T}$}}}}({\bf x},{\bf x}_F)$ (from equation (\ref{eq60b})).
Versions of the Marchenko method based on expressions similar to equations (\ref{eq13genGdeco}) and (\ref{eq13genGdecoup}) have already been implemented for
the retrieval of decomposed elastodynamic Green's functions in lossless media, ignoring evanescent waves \citep{Wapenaar2014GJI, Costa2014PRE, Reinicke2019WM, Reinicke2020GEO}.  

\section{Symmetry properties of the propagator and decomposition matrices}\label{AppC}

Let $N\times N$ propagator matrices 
${\bf W}({\bf x},{\bf x}_A)$ and ${\bf W}({\bf x},{\bf x}_B)$ be two independent solutions of equation (\ref{eq2.1}) (with unified operator matrix ${{\mbox{\boldmath ${\cal A}$}}}({\bf x})$
defined in equation (\ref{eq7em})), 
with boundary condition (\ref{eq9998d}), modified for coordinate vectors ${\bf x}_A$ and ${\bf x}_B$. We show that the quantity 
$\int_{{\mathbb{R}}^2}{\bf W}^t({\bf x}_{\rm H},x_3,{\bf x}_A){\bf N}{\bf W}({\bf x}_{\rm H},x_3,{\bf x}_B){\rm d}^2{\bf x}_{\rm H}$
is a `propagation invariant', meaning that it is independent of $x_3$ \citep{Haines88GJI, Kennett90GJI, Koketsu91GJI, Takenaka93WM}.
To this end we take the derivative in the $x_3$-direction, apply the product rule for differentiation, use equation (\ref{eq2.1}) and symmetry relation (\ref{eqsym1f}), according to
\begin{eqnarray}
\partial_3\int_{{\mathbb{R}}^2}{\bf W}^t({\bf x},{\bf x}_A){\bf N}{\bf W}({\bf x},{\bf x}_B){\rm d}^2{\bf x}_{\rm H}
=\int_{{\mathbb{R}}^2}{\bf W}^t({\bf x},{\bf x}_A)\bigl(\underbrace{\mbox{\boldmath ${\cal A}$}^t{\bf N}+{\bf N}\mbox{\boldmath ${\cal A}$}}_{\bf O}\bigr){\bf W}({\bf x},{\bf x}_B){\rm d}^2{\bf x}_{\rm H}={\bf O}.
\end{eqnarray}
This confirms that the integral is a propagation invariant. In a similar way, using symmetry relation (\ref{eqsym2f}), it can be shown that 
$\int_{{\mathbb{R}}^2}\bar{\bf W}^\dagger({\bf x}_{\rm H},x_3,{\bf x}_A){\bf K}{\bf W}({\bf x}_{\rm H},x_3,{\bf x}_B){\rm d}^2{\bf x}_{\rm H}$ is also a propagation invariant.
Using boundary condition (\ref{eq9998d}), modified for $x_3=x_{3,A}$ and $x_3=x_{3,B}$, we find from the first propagation invariant
\begin{eqnarray}\label{eq65aw}
{\bf W}^t({\bf x}_B,{\bf x}_A)={\bf N}{\bf W}({\bf x}_A,{\bf x}_B){\bf N}^{-1}
\end{eqnarray}
and from the second propagation invariant
\begin{eqnarray}\label{eq65awk}
\bar{\bf W}^\dagger({\bf x}_B,{\bf x}_A)={\bf K}{\bf W}({\bf x}_A,{\bf x}_B){\bf K}^{-1}.
\end{eqnarray}
From equations (\ref{eq65aw}) and (\ref{eq65awk}), using ${\bf K}{\bf N}^{-1}={\bf J}$, we find
\begin{eqnarray}\label{eq65aws}
\bar{\bf W}({\bf x}_B,{\bf x}_A)={\bf J}{\bf W}^*({\bf x}_B,{\bf x}_A){\bf J}^{-1}.
\end{eqnarray}
This equation is used in section \ref{sec2.3} and Appendix \ref{sec4.3} in the derivation of the relation between the propagator matrix and the Marchenko focusing functions.

To derive a symmetry property for $\tilde{\bf D}_1^\pm=\tilde{\bf L}_2^\pm (\tilde{\bf L}_1^\pm)^{-1}$ (equation (\ref{eq2.14rev})), we start by
Fourier transforming symmetry relations (\ref{eqsym1f}) -- (\ref{eqsym3f}), assuming the medium is laterally invariant at the depth level where the transformation is applied. This gives
\begin{eqnarray}
\tilde{\bf A}^t(-{{\bf s}},x_3)&=&-{\bf N}\tilde{\bf A}({{\bf s}},x_3){\bf N}^{-1},\label{eqsymB}\\
\tilde{\bf A}^\dagger({{\bf s}},x_3)&=&-{\bf K}{\tilde{\bar{\bf A}}}({{\bf s}},x_3){\bf K}^{-1},\label{eqsymC}\\
\tilde{\bf A}^*(-{{\bf s}},x_3)&=&{\bf J}{\tilde{\bar{\bf A}}}({{\bf s}},x_3){\bf J}^{-1}.\label{eqsymD}
\end{eqnarray}
The eigenvalue decomposition of matrix $\tilde{\bf A}$ is defined in equation (\ref{eqA21}),
with the partitioning of $\tilde{\bf \Lambda}$ and $\tilde{\bf L}$ defined in equation (\ref{Aeq7mvbbprffBBrev}).
For all wave phenomena mentioned in Appendix \ref{sec4.1}, 
the eigenvalue matrix $\tilde{{{\mbox{\boldmath $\Lambda$}}}}$ obeys the following symmetry relations
\begin{eqnarray}
\tilde{{{\mbox{\boldmath $\Lambda$}}}}^t(-{\bf s},x_3)&=&-{\bf N}\tilde{{{\mbox{\boldmath $\Lambda$}}}}({\bf s},x_3){\bf N}^{-1},\label{eqB5}\\
\tilde{{{\mbox{\boldmath $\Lambda$}}}}^\dagger({\bf s},x_3)&=&-{\bf J}\tilde{\bar{{\mbox{\boldmath $\Lambda$}}}}({\bf s},x_3){\bf J}^{-1}.\label{eqB6}
\end{eqnarray}
Given equations (\ref{eqA21}) and (\ref{eqsymB}) -- (\ref{eqB6}), matrix $\tilde{\bf L}$ can be normalized such that
\begin{eqnarray}
\tilde{\bf L}^t(-{\bf s},x_3)&=&-{\bf N}\tilde{\bf L}^{-1}({\bf s},x_3){\bf N}^{-1},\label{eqB9}\\
\tilde{\bf L}^\dagger({\bf s},x_3)&=&{\bf J}{\tilde{\bar{\bf L}}}^{-1}({\bf s},x_3){\bf K}^{-1}.\label{eqB10}
\end{eqnarray}
From the latter two equations, we obtain
\begin{eqnarray}
\tilde{\bf L}^*({\bf s},x_3)&=&{\bf J}{\tilde{\bar{\bf L}}}(-{\bf s},x_3){\bf K},\label{eqB11}
\end{eqnarray}
or, using the partitioning of $\tilde{\bf L}$ as defined in equation (\ref{Aeq7mvbbprffBBrev}),
\begin{eqnarray}
\{\tilde{\bf L}_1^\pm({\bf s},x_3)\}^*&=&{\bf J}_{11}{\tilde{\bar{\bf L}}}_1^\mp(-{\bf s},x_3){\bf K}_{12},\label{eqB12}\\
\{\tilde{\bf L}_2^\pm({\bf s},x_3)\}^*&=&{\bf J}_{22}{\tilde{\bar{\bf L}}}_2^\mp(-{\bf s},x_3){\bf K}_{12},\label{eqB13}
\end{eqnarray}
with ${\bf J}_{11}$ and ${\bf J}_{22}$ being the upper-left and lower-right submatrices of matrix ${\bf J}$, and ${\bf K}_{12}$ being the upper-right (= lower-left) submatrix of matrix ${\bf K}$.
From equations (\ref{eqB12}) and (\ref{eqB13}) it follows that $\tilde{\bf D}_1^\pm$ as defined in equation (\ref{eq2.14rev}) obeys the following symmetry relation
\begin{eqnarray}\label{eq34B}
{\tilde{\bar{\bf D}}}_1^\pm({\bf s},x_3)={\bf J}_{22}\{\tilde{\bf D}_1^\mp(-{\bf s},x_3)\}^*{\bf J}_{11}^{-1}.
\end{eqnarray}
Finally, using $\tilde{\bf \Delta}_1=\tilde{\bf D}_1^+- \tilde{\bf D}_1^-$ (equation \ref{eq2107}), we obtain
\begin{eqnarray}
\{{\tilde{\bar{\bf \Delta}}}_1({\bf s},x_3)\}^{-1}=-{\bf J}_{11}\{{\tilde{{\bf \Delta}}}_1^*(-{\bf s},x_3)\}^{-1}{\bf J}_{22}^{-1}.\label{eqC16}
\end{eqnarray}

\newpage
\centerline{\Huge Captions}
\mbox{}\\

\noindent
Fig 1. Relations between the propagator matrix ${\bf W}({\bf x},{\bf x}_F)$, the transfer matrix ${{{\mbox{\boldmath ${\cal T}$}}}}({\bf x},{\bf x}_F)$, 
and the Marchenko focusing functions $F^p({\bf x},{\bf x}_F)$ and
$F^v({\bf x},{\bf x}_F)$ (right column of ${\bf Y}({\bf x},{\bf x}_F)$). 
The green and yellow double-sided arrows indicate full wave fields (implicitly consisting of downgoing and upgoing components), whereas the red and blue
single-sided arrows indicate decomposed downgoing and upgoing wave fields, respectively.
\\

\noindent
Fig 2. (a) Horizontally layered medium.  (b) Propagator matrix component $W^{p,p}(s_1,x_3,x_{3,F},\tau)$ (for fixed $s_1=1/3500$ m/s).
(c) Propagator matrix component $W^{p,v}(s_1,x_3,x_{3,F},\tau)$. 
\\

\noindent
Fig 3. (a) Focusing function $F^p(s_1,x_3,x_{3,F},\tau)$ (for fixed $s_1=1/3500$ m/s).
(b) Time-reversed focusing function $F^p(s_1,x_3,x_{3,F},-\tau)$. 
\\

\noindent
Fig 4. (a) Decomposed focusing function $F^-(s_1,x_3,x_{3,F},\tau)$ (for fixed $s_1=1/3500$ m/s).
(b) Decomposed focusing function $F^+(s_1,x_3,x_{3,F},\tau)$ .
\\

\end{spacing}
\end{document}